\documentclass[pmlr]{jmlr}

\usepackage{longtable}
\usepackage{multirow}
\usepackage{tablefootnote}
\usepackage[font=small,skip=15pt]{caption}

\usepackage{booktabs}
\usepackage[load-configurations=version-1]{siunitx} 
\usepackage{siunitx}
\usepackage{enumitem}

\makeatletter
\def\set@curr@file#1{\def\@curr@file{#1}} 
\makeatother

\theorembodyfont{\upshape}
\theoremheaderfont{\scshape}
\theorempostheader{:}
\theoremsep{\newline}

\jmlryear{2020}
\jmlrworkshop{Machine Learning for Healthcare}

\title[Towards data-driven stroke rehabilitation via wearable sensors and deep learning]{Towards data-driven stroke rehabilitation\\ via wearable sensors and deep learning}

\author{\Name{Aakash Kaku}\thanks{Equal contribution} \Email{ark576@nyu.edu} 
      \addr \\Center for Data Science\\
      New York University\\
      \AND
      \Name{Avinash Parnandi}$^*$ \Email{avinash.Parnandi@nyulangone.org}
        \addr \\Department of Neurology\\
      New York University School of Medicine\\
      \AND
      \Name{Anita Venkatesan} \Email{anita.venkatesan@nyulangone.org}
        \addr \\Department of Neurology\\
      New York University School of Medicine\\
      \AND
      \Name{Natasha Pandit} \Email{ngp238@nyu.edu}
        \addr \\Department of Neurology\\
      New York University School of Medicine\\
      \AND
      \Name{Heidi Schambra}\thanks{Joint corresponding/last authors.} \Email{Heidi.Schambra@nyulangone.org } 
         \addr \\Department of Neurology\\
      New York University School of Medicine\\
      \AND
      \Name{Carlos Fernandez-Granda}$^\dagger$ \Email{cfgranda@cims.nyu.edu } 
        \addr \\Center for Data Science\\ Courant Institute of Mathematical Sciences \\
      New York University\\
        } 

\begin{document}

\maketitle

\begin{abstract}
  Recovery after stroke is often incomplete, but rehabilitation training may potentiate recovery by engaging endogenous neuroplasticity. In preclinical models of stroke, high doses of rehabilitation training are required to restore functional movement to the affected limbs of animals. In humans, however, the necessary dose of training to potentiate recovery is not known. This ignorance stems from the lack of objective, pragmatic approaches for measuring training doses in rehabilitation activities. Here, to develop a measurement approach, we took the critical first step of automatically identifying functional primitives, the basic building block of activities. Forty-eight individuals with chronic stroke performed a variety of rehabilitation activities while wearing inertial measurement units (IMUs) to capture upper body motion. Primitives were identified by human labelers, who labeled and segmented the associated IMU data. We performed automatic classification of these primitives using machine learning. We designed a convolutional neural network model that outperformed existing methods. The model includes an initial module to compute separate embeddings of different physical quantities in the sensor data. In addition, it replaces batch normalization (which performs normalization based on statistics computed from the training data) with instance normalization (which uses statistics computed from the test data). This increases robustness to possible distributional shifts when applying the method to new patients. With this approach, we attained an average classification accuracy of 70\%. Thus, using a combination of IMU-based motion capture and deep learning, we were able to identify primitives automatically. This approach builds towards objectively-measured rehabilitation training, enabling the identification and counting of functional primitives that accrues to a training dose.
  
\end{abstract}

\section{Introduction} \label{sec:intro}
Stroke is the leading cause of disability in the United States, affecting nearly 1 million individuals annually and costing the US an estimated \$240 billion \citep{intro1,intro2}. Almost two-thirds of stroke patients have significant motor impairment in their upper extremities (UE), which  limits their performance of activities of daily living (ADLs) like feeding, bathing, grooming, and dressing. Rehabilitation training, incorporating the repeated practice of ADLs, is the primary clinical intervention to reduce UE impairment. However, rehabilitation is increasingly believed to have a marginal impact on recovery because of its low numbers of functional repetitions, or training dose \citep{krakauer2012getting}. In animals models, UE recovery is substantially improved by high-dose functional training delivered early after stroke \citep{murata2008effects,jeffers2018does}. In humans, the optimal training dose to improve recovery is unknown, because no quantitative dose-response studies have been undertaken in the early weeks after stroke. The resulting vacuum of clinical guidelines has perpetuated the delivery of low and variable training doses \citep{lang2009observation}.

A major reason for this failure is the absence of precise and pragmatic tools to measure training dose. Most rehabilitation studies use time-in-therapy to approximate dose \citep{lohse2018reporting}. Although one may intuit that more scheduled time equals more training repetitions, a linear relationship does not hold. In a seminal study observing standard rehabilitation practice, investigators found that the number of trained movements varied widely across clinicians and sessions \citep{lang2009observation}, underscoring the imprecision of using time-in-therapy as a proxy for dose. Another approach for measuring dose is manual tallying, where a human observer identifies and counts motions of interest. Because functional motions are fluid and fast, they are difficult to disambiguate in real time. Video recordings aid scrutiny, but analysis is prohibitively time-intensive: in our experience, one minute of videotaped motion requires one hour of analysis by trained coders. This laboriousness makes manual tallying impractical for clinical or research deployment. 

A third approach for measuring training dose is pairing motion capture technology with machine learning. Wearable devices such as inertial measurement units (IMUs) generate kinematic data about UE motions. Investigators decide on motions of interest (classes) that they wish to detect. Using a supervised approach, machine learning models can be trained to recognize classes of motions from their kinematic signatures \citep{parnandi2019pragmatic}. Once these motions are detected, they can be tallied to a dose. 

Recent studies using this approach have sought to classify functional motion (e.g. tying shoelaces) and nonfunctional motion (e.g. arm swinging during walking) \citep{mcleod2016using,bochniewicz2017measuring,leuenberger2017method}. In one, chronic stroke patients performed loosely-structured activities while wearing an IMU on their paretic wrist \citep{bochniewicz2017measuring}. From the IMU recordings, a random-forest model distinguished functional from nonfunctional motion with 70\% accuracy. The resulting unit of measure was time spent in functional motion. While the classification performance of this approach is good, the resulting metric is nearly as problematic as measuring time-in-therapy: for example, did more time in functional motion correspond to the performance of more motions, or did it simply take longer to perform the same motions? What kinds of functional motions were made? Without knowing motion content, it is challenging to identify the relationship between repetitions and recovery, or to replicate a successful rehabilitation intervention.

In this work, we sought to address these limitations by taking the first step towards measuring rehabilitation dose. To unpack the motion content of rehabilitation, we focus on functional primitives, single motions or minimal-motions that serve a single purpose \citep{schambra2019taxonomy}. There are five classes of functional primitives: reach (motion to contact an object), transport (motion to convey an object), reposition (motion into proximity of an object), stabilize (minimal-motion to keep an object still), and idle (minimal-motion to stand at the ready). Rehabilitation activities can be successfully broken down into these constituent primitives, indicating that primitives are a useful unit of measure \citep{schambra2019taxonomy}. As a unit of measure, primitives thus provide motion content information that would inform a dose-response inquiry and the replication of an intervention. We further focus on primitives for three reasons. First, because primitives are a single motion event with a surprisingly consistent phenotype, even in stroke patients \citep{schambra2019taxonomy}, automated identification is facilitated. Second, because some stroke patients are unable to fully complete activities, primitives can provide a more nuanced picture of performance. Third, because primitives may be neurally hard-wired \citep{graziano2016ethological,ramanathan2006form}, measuring their execution may enable us to more precisely track central nervous system reorganization after stroke. 

\begin{figure}[t]
  \centering 
  \includegraphics[width = \linewidth]{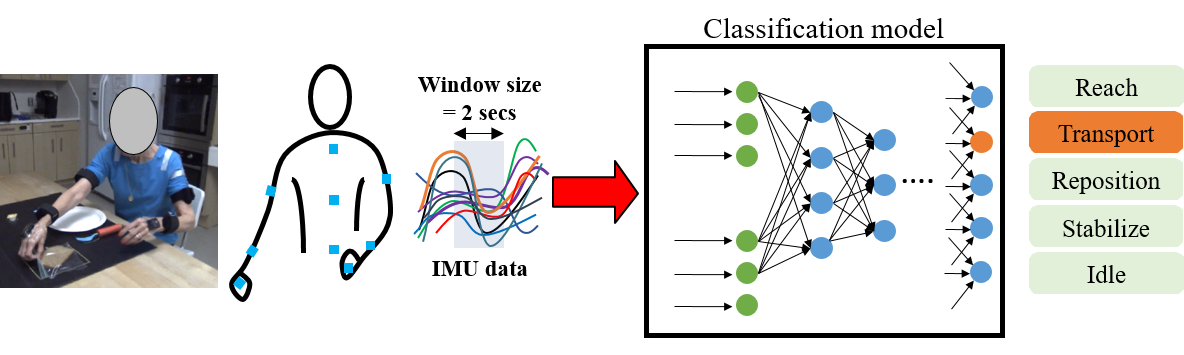} 
  \caption{Diagram of the proposed approach for identification of functional primitives. Stroke patients perform a battery of rehabilitation activities while wearing IMU sensors. Machine-learning models are trained to classify the functional primitives from the sensor data.}
  \label{fig:schematic_problem_overview} 
\end{figure}

To develop an approach that identifies and counts functional primitives in a practical, automated manner, we paired sensor-based motion capture with supervised machine learning. We used an array of inertial measurement units (IMUs) on the upper body to generate richly characterized motion data. We had stroke patients perform a battery of rehabilitation activities, which generated a large sample of primitives with varying characteristics (e.g. speed, duration, extent, location in space). Once the motion data was labeled, we trained various machine learning models to classify primitives. We report our steps for identifying the best-performing algorithm and for optimizing its classification performance. Our approach is illustrated in Figure~\ref{fig:schematic_problem_overview}.

\paragraph{Generalizable Insights about Machine Learning in the Context of Healthcare}
In this work, we performed a systematic comparison of machine learning methods for the task of functional-primitive identification, and propose a model that outperforms existing methodology. Our results suggest several insights that have the potential to generalize to other healthcare applications (especially those involving wearable sensors). First, deep learning methods that directly process multivariate time series of sensor data seem to be significantly more effective than techniques based on handcrafted statistical features. Second, in order to combine data that represents different physical quantities, it may be helpful to map them to a common representation space incorporating an initial module that produces a separate embedding for each quantity. Third, adaptive feature-normalization techniques, such as instance normalization, may increase the robustness of convolutional neural networks to shifts in the distribution of the data, which can occur when the models are applied to new patients. Adaptive normalization uses statistics computed on the test data, in contrast to batch normalization, which uses statistics computed on the training data. 

\section{Related Work} \label{sec:relevant_work}
To the best of our knowledge, only one previous study has used machine learning to identify functional primitives from IMU sensor data \citep{heidi}. The authors used hidden Markov models to learn a latent representation of the sensor data, which was then used to perform classification via logistic regression. They acquired their data based on a few highly structured tasks, primarily consisting of moving objects to/from horizontal and vertical targets. Although this approach is useful to develop proof-of-concept methods, it does not reflect many of the challenges of real-world scenarios where unstructured tasks generate more varied and complex motions. In the present work, we gather data from real-world rehabilitation activities. Modeling functional primitives in this setting requires more complex models such as deep neural networks. In addition, the previous work was based on a small number of mostly mildly impaired patients; the present work increases the sample size 8-fold and captures a wider range of impairment. 

Activity recognition using data gathered with wearable sensors is an active area of research in machine learning. However, it is important to emphasize that recognizing activities does not address the problem of measuring rehabilitation dose. Activities are prolonged sequences of motions that achieve several goals \citep{schambra2019taxonomy}. Problematically, activities are not standardized: their motion content varies by individual, culture, and environment \citep{fisher1992cross,teresi1989some}. For example, the motions undertaken to perform a cooking activity differ if the meal is breakfast or dinner, or Japanese or German. This variable motion content not only challenges the automated recognition of activities, it also limits the identification of a dose-response relationship and the reproducibility of interventions.  

Although activity recognition does not serve dose quantitation, prior studies in this area offer computational directions for classifying patterns of motion. Initially, methodology was mostly based on statistical features processed with techniques such as random forests or fully-connected neural networks (e.g. \cite{stat1,kwapisz2011activity}). More recently, deep learning methods have been applied to perform activity recognition without precomputing statistical features. Specifically, \cite{wang2017time} showed that a ResNet-style convolutional architecture outperformed traditional non-deep learning methods as well as fully convolutional networks on several  activity-recognition datasets \citep{macro_action_1, macro_action_2, macro_action_3}. \cite{cui2016multi} demonstrated that a simple convolutional model performed well when trained on data sampled at multiple scales. \cite{nn_1}, \cite{nn_2} and \cite{murad2017deep} successfully used recurrent networks like Long Short Term Memory (LSTM) and Bi-LSTM for activity recognition. However, \cite{ha2015multi} found that convolutional neural networks may outperform recurrent networks for some tasks. Given these conflicting results, in this work we sought to determine the necessity of using statistical features and the performance of recurrent versus convolutional networks for classification of functional primitives.

\section{Cohort} \label{sec:cohort}
\subsection{Cohort Selection}
We collected motion data from 48 stroke patients in an inpatient rehabilitation setting. Individuals were included if they were $\geq$ 18 years old, had premorbid right-handed dominance, and had unilateral weakness from either ischemic or hemorrhagic stroke. Individuals were excluded if they had traumatic brain injury; any musculoskeletal or non-stroke neurological condition that interferes with the assessment of motor function; contracture at the shoulder, elbow, or wrist;  moderate upper extremity dysmetria or truncal ataxia; visuospatial neglect; apraxia; global inattention; or legal blindness. Table~\ref{tab:demographics-table} describes the demographic and clinical characteristics of the patients.

\begingroup
\renewcommand*{\arraystretch}{1.1}
\begin{table}[t]
\centering
\begin{center}
\begin{tabular}{|c | c | c | c |}
\hline
 & Training set & Test set 1 & Test set 2 \\
\hline
n & 33 & 8 & 7\\
\hline
Age (in years) & 56.3 (21.3-84.3) & 60.9 (42.6-84.3) & 58.3 (41.1-74.4)\\
\hline
Gender & 18 F : 15 M & 4 F : 4 M & 4 F: 3 M\\
\hline
Time since stroke (in years) & 6.5 (0.3-38.4) & 3.1 (0.4-5.7) & 3.16 (1.1-6.4)\\
\hline
Paretic side (Left : Right) & 18 L : 15 R & 4 L : 4 R & 3 L : 4 R\\
\hline
Stroke type & & & \\
(Ischemic : Hemorrhagic) & 30 I : 3 H & 8 I : 0 H & 2 I : 5 H\\
\hline
Fugl-Meyer Assessment score & 48.1 (26-65)& 49.4 (27-63) & 15.3 (8-23)\\
\hline
\end{tabular}
\end{center}
\caption{Demographic and clinical characteristics of the patients in the cohort. Mean and ranges in parentheses are shown. The cohort is divided into a training set and a test set (Test set 1) of mildly and moderately-impaired patients, and a test set of severely-impaired patients (Test set 2). There is no overlap of patients between the training and test sets. 
}
\label{tab:demographics-table}
\end{table}
\endgroup

\subsection{Data Acquisition and Labelling}
\label{sec:data}
The data were gathered while the patients performed activities of daily living that are commonly trained during stroke rehabilitation. The activities included: washing the face, applying deodorant, combing the hair, donning and doffing glasses, preparing and eating a slice of bread, pouring and drinking a cup of water, brushing teeth, and moving an object on horizontal and vertical target array. See Section~\ref{sec:app_desc_act} for a detailed description. The patients performed five repetitions of each activity.

Upper extremity motion was recorded using nine IMUs (Noraxon) attached to the upper body, specifically to the cervical vertebra C7, the thoracic vertebra T12, the pelvis, and both arms, forearms, and hands. Each IMU samples linear acceleration, angular velocity, and magnetic heading at 100 Hz. These data are then converted to 9 sensor-centric unit quaternions, representing the rotation of each sensor on its own axes, using coordinate transformation matrices. In addition, proprietary software (Myomotion, Noraxon) generates 22 anatomical angle values using a rigid-body skeletal model scaled to the patient's height and UE segment lengths. See Section~\ref{sec:app_joint_angles} for a detailed description of these angles. This results in a 76-dimensional vector containing the linear acceleration, quaternion, and joint-angle information. As additional features, we included the time elapsed from the start of the activity in seconds and the paretic side of the patient (left or right) encoded in a one-hot vector. This increases the dimension of the feature vector to 78. Each entry (except the one indicating the paretic side) was mean-centered and normalized separately for each task repetition in order to remove spurious offsets introduced during sensor calibration. 

In order to label the data, motion was synchronously captured using two cameras (1088 x 704, 60 frames per second; Ninox, Noraxon) placed orthogonally $<$ 2 m from the patient. Trained observers watched the videos to identify and label functional primitives in the video, which simultaneously labeled primitives in the IMU data.

\subsection{Evaluation Protocol} \label{sec:vali_protocol}

An important consideration when evaluating methodology for classification of functional primitives is the level of impairment of the patients. Impairment level was assessed using the upper extremity Fugl-Meyer Assessment (FMA), where a higher score indicates less impairment (the maximum score is 66) \citep{fm}. We separated the patients into three levels of impairment according to their FMA score: mild (FMA  53-65), moderate (FMA 26-52), and severe (FMA 0-25) \citep{woodbury2013rasch}. In order to evaluate our methodology, we assigned the patients to a training set containing 33 mildly and moderately patients, a test set containing 8 mildly and moderately impaired patients (Test set 1)\footnote{The 41 mildly and moderately patients were separated into eight subgroups, balancing for impairment level and their paretic side (left or right). One patient in each group was randomly assigned to the test set. The remaining patients were assigned to the training set.}, and an additional test set containing 7 severely impaired patients (Test set 2). Table~\ref{tab:demographics-table} describes the characteristics of these datasets. Our first goal was to test the methods on patients with a similar impairment level as those used for training. Our second goal was to evaluate the generalizability of the trained model to patients with worse impairment. To avoid any selection bias or data leakage, the training and test sets were constructed before training any models, and model selection was carried out via cross-validation based exclusively on the training set (see Section~\ref{sec:experiments} for more details). 


\section{Methodology} \label{sec:methodology}

Our goal in this work was to design a machine learning model for the identification of functional primitives from sensor data. We framed this as a classification problem, where the input to the model was a window of the multidimensional time series obtained from the IMU sensors (see Section~\ref{sec:data} for a detailed description), and the output was an estimate of the primitive corresponding to the center of the window. Note, however, that a significant portion of the window could contain motion corresponding to other functional primitives. The duration of the window was set to 2 seconds in order to provide sufficient context to the model, i.e. from the time steps flanking the center of the window (shorter windows yielded inferior results in preliminary experiments). In this section, we describe two key modifications to standard convolutional neural network architectures: learned embeddings that map each sensor to a common representation, and adaptive normalization of the network features. These modifications yield a model that outperformed existing techniques for primitive identification, as demonstrated by the results reported in Section~\ref{sec:results}.

\subsection{Learning Embeddings for Diverse Inputs} \label{sec:input_embeddings}
\begin{figure}[t]
  \centering 
  \includegraphics[width = \linewidth]{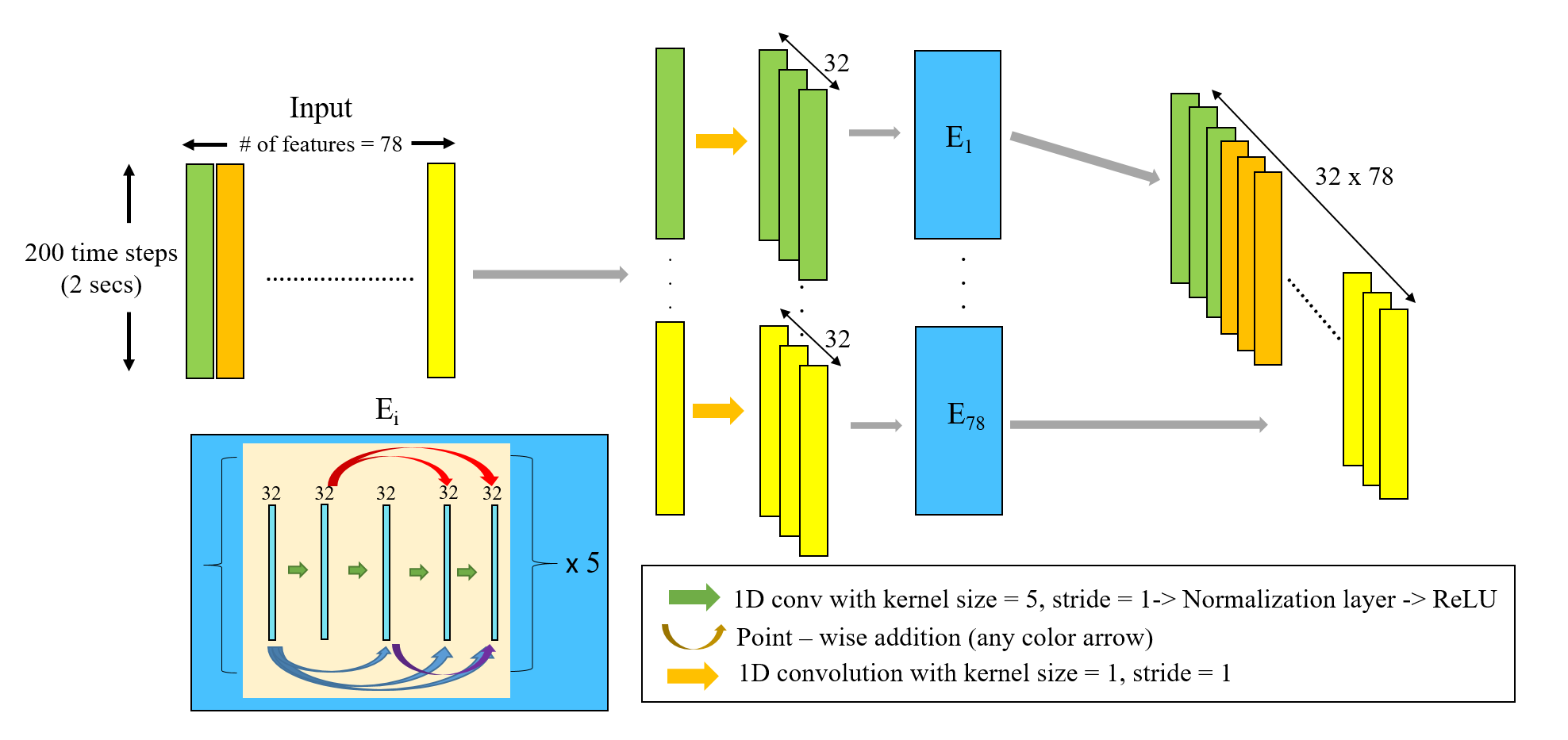} 
  \caption{Diagram of the proposed approach to process multivariate time series where each entry may represent a different physical quantities. The $i$th entry is fed to an embedding module (denoted by $E_i$) consisting of several blocks of convolutional layers with DenseNet-like connections~\citep{huang2017densely}. The weights of each module are not shared, so that they can be calibrated to adapt to the corresponding physical quantity.}
  \label{fig:diff_emb_n_emb} 
\end{figure}

Each layer in a convolutional neural network (CNN) computes local linear combinations of outputs of the previous layer, weighted by the coefficients of several convolutional filters. As a result, when we apply a CNN to the multivariate time series representing the IMU data, the different entries in the time series are combined at the second layer. This may be problematic because each entry represents very different kinematic information, such as accelerations, quaternions, and joint angles. To address this issue, we mapped each entry separately to a common representation space. The mapping was implemented using multiple embedding modules consisting of several convolutional layers. Each embedding module processes one of the entries in the time series. The embeddings were then concatenated and fed to a CNN. The embedding modules were optimized jointly with the CNN. Figure~\ref{fig:diff_emb_n_emb} shows a diagram of our proposed approach. A related previous work by \cite{yao2017deepsense} proposed computing embeddings in the frequency domain. 


\subsection{Robust Generalization via Adaptive Feature Normalization}

In order to develop models that can be deployed in realistic rehabilitation settings, it is critical to ensure that they generalize accurately \emph{to new patients} not present in the training set. This is challenging due to varying impairment levels and movement idiosyncrasies, which may produce systematic differences between the training data and the data from new patients. Achieving robustness to systematic shifts between the training and test data is a fundamental challenge in modern machine learning, particularly in healthcare applications. In the case of CNNs, recent work by~\cite{kaku2020like} suggests that batch normalization may be particularly sensitive to such shifts.

Batch normalization has become a standard element in CNNs because it provides stability to different initializations and learning rates~\citep{ioffe2015batch}. It consists of two operations applied at the end of each layer. First, the features corresponding to each convolutional filter in the layer are centered and normalized using an approximation to their mean and standard deviation. Second, the resulting normalized features are scaled and shifted using two learned parameters per filter (a scaling factor and a shift). When the CNN is being trained, the estimates of the mean and standard deviation are obtained by averaging over each batch of examples. Simultaneously, estimates of the population mean and standard deviation of each filter are computed via running averages. The population statistics are used to perform normalization at test time. However, if the distributions of the training and test data differ, then these statistics may not center and normalize the data adequately, as demonstrated by~\cite{kaku2020like}. 

Following~\cite{kaku2020like}, we applied CNN models to perform identification of functional primitives by replacing batch normalization with instance normalization, a normalization technique originally proposed to promote style invariance in image-style transfer by \cite{ulyanov2016instance}. In instance normalization, the features for each convolutional filter are centered and normalized using means and standard deviations that are computed over each individual example both at training and test time. This avoids the mismatch of training and test statistics that may occur with batch normalization. 

\section{Computational Experiments} \label{sec:experiments}
The goal of our computational experiments was to compare the performance of different machine learning methods for identification of functional primitives, and to test our proposed approach. Inspired by the existing literature on movement identification from sensor data, we applied techniques based on statistical features (random forests and fully-connected neural networks), convolutional neural networks, and recurrent neural networks. As explained in Section~\ref{sec:methodology}, we framed primitive identification as a classification problem, where time-series windows were assigned to five different classes. We carried out model selection and hyperparameter optimization via cross-validation exclusively on the training set of 33 patients (see Section~\ref{sec:vali_protocol}). To this end, we performed four different random splits. Each split contained 24 or 25 patients for training, and 9 or 8 patients for validation. Each patient appears in exactly one validation set. During validation, the models were compared using average classification accuracy\footnote{To be clear, if $c$ out of a $n$ windows are classified correctly, the classification accuracy equals $c/n$.} across the four splits. Section~\ref{sec:additional_results} reports the validation results. For each of the methods described below, we selected the hyperparameters achieving the highest cross-validation accuracy. Then, fixing those hyperparameters, we evaluated an ensemble of the models corresponding to the different splits on the two test sets. The ensemble was computed by averaging the estimated probabilities produced by each model (this resulted in a small improvement in accuracy with respect to the validation results for all methods).

In the remainder of this section we describe the hyperparameters of the different machine-learning methods in more detail. All neural-network models were trained using the Adam optimizer~\citep{kingma2014adam} with starting learning rate of 1.25 $10^{-4}$, which was divided by two every 20 epochs for the fully-connected and convolutional networks, and every 10 epochs for the recurrent networks. Training was terminated via early stopping based on the validation accuracy.

{\setlength{\parindent}{0cm}
\textbf{Random forest}: The input to the random forest models was a set of five statistics computed over each dimension of the 78-dimensional windows: mean, maximum, minimum, standard deviation, and root mean square. These features capture useful information for movement identification, such as the energy of the motion and the variations within the window~\citep{kwapisz2011activity,heidi}. We used the scikit-learn random forest implementation~\citep{scikit-learn}.\\
\emph{Hyperparameters}: Minimum number of examples required to split each internal node, and minimum number of samples required to be at a leaf. The selected values were 2 and 1 respectively.
}

{\setlength{\parindent}{0cm}
\textbf{Fully-connected neural network}: The input to the fully-connected neural network was the same set of five statistics as for the random forest.\\
\emph{Hyperparameters}: Number of layers, number of neurons per layer, and dropout rate. The selected values were 8, 900, and 0.5 respectively.
}

{\setlength{\parindent}{0cm}
\textbf{Recurrent neural network}: We used one of the most popular recurrent architectures, Long Short Term Memory (LSTM). Preliminary experiments with a Bi-LSTM architecture yielded inferior performance. The LSTM received the windows of the multivariate time series directly as an input.\\
\emph{Hyperparameters}: Dimensionality of hidden units. The selected value was 4000.
}

{\setlength{\parindent}{0cm}
\textbf{Convolutional neural network (CNN)}: As in the case of the LSTM, we used CNNs to process the time-series window directly. We chose two architectures with skip connections similar to the ResNet~\citep{wang2017time} and the DenseNet~\citep{huang2017densely}. Preliminary experiments with an AlexNet-style architecture \citep{le2016data} yielded worse performance. In order to evaluate the effect of input embeddings and adaptive feature normalization (see Section~\ref{sec:methodology}), we performed an ablation analysis where we trained the four possible combinations of these design choices for each model (with/without input embedding, with batch normalization/instance normalization). The depth of all networks was set to 44 layers. The architectures are described in detail in Section~\ref{sec:app_schematic_conv}. }

\begingroup
\renewcommand*{\arraystretch}{1.25}
\begin{table}[t]
\centering
\begin{tabular}{|c|c|c|c|c|c|c|}
\hline
\multicolumn{7}{|c|}{Mildly / Moderately-impaired patients (Test set 1)} \\
 \cline{1-7}
\begin{tabular}[c]{@{}c@{}}Method \end{tabular} & \begin{tabular}[c]{@{}c@{}}Random  forest\end{tabular} & FCNN & \begin{tabular}[c]{@{}c@{}}CNN\end{tabular} & LSTM & \begin{tabular}[c]{@{}c@{}}Proposed \end{tabular} & Ensemble \\ \hline
\begin{tabular}[c]{@{}c@{}}Balanced\\ accuracy\end{tabular} & 52.98 & 58.04 & 64.01 & 66.58 & 69.21 & \textbf{70.11} \\ \hline
\multicolumn{7}{|c|}{Severely-impaired patients (Test set 2)} \\
 \cline{1-7}
\begin{tabular}[c]{@{}c@{}}Method \end{tabular} & \begin{tabular}[c]{@{}c@{}}Random forest\end{tabular} & FCNN & \begin{tabular}[c]{@{}c@{}}CNN\end{tabular} & LSTM & \begin{tabular}[c]{@{}c@{}}Proposed \end{tabular} & Ensemble \\ \hline
\begin{tabular}[c]{@{}c@{}}Balanced\\ accuracy\end{tabular} & 32.95 & 36.60 & 38.22 & 41.76 & 43.50 & \textbf{44.39} \\ \hline
\end{tabular}
\caption{Balanced accuracy on Test set 1 and Test set 2 of the machine-learning models described in Section \ref{sec:experiments}. FCNN denotes fully connected neural network. The ensemble is a combination of the proposed model and the LSTM, where the output probabilities were averaged.}
\label{tab:final_results_table}
\end{table}
\endgroup

\section{Results} \label{sec:results}
\begin{figure}[t]
  \centering 
  \includegraphics[width=0.75\linewidth]{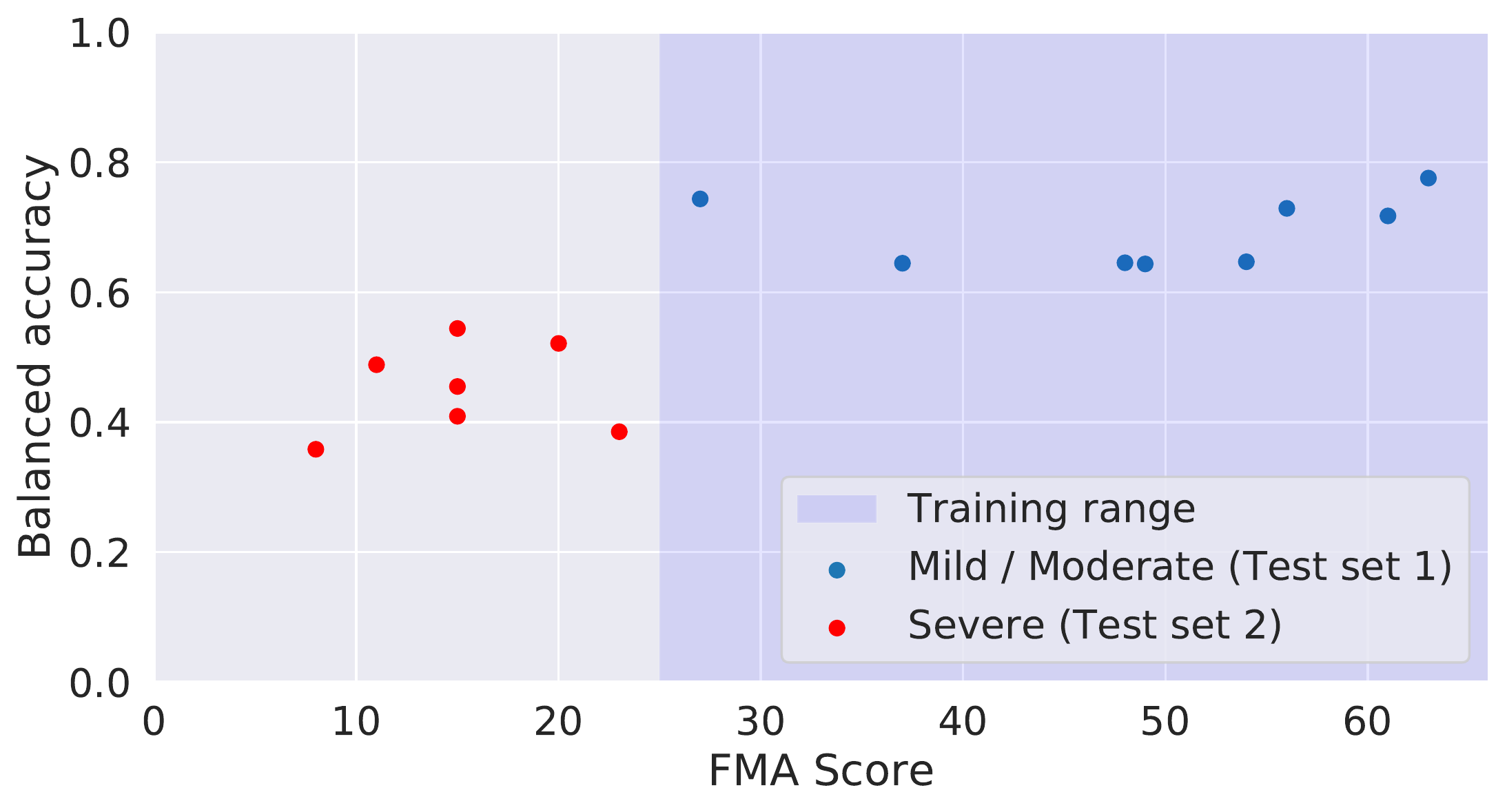} 
  \caption{Balanced accuracy of the ensemble model applied to the patients in Test set 1 and Test set 2, plotted as a function of their impairment level (quantified by FMA score). The range of impairment levels of the patients in the training set is indicated by the dark-colored background. The balanced accuracy was above 60\% for all mildly/moderately impaired patients, and above 50\% for three out of the seven severely impaired patients (for comparison, random assignment yields 20\% accuracy).}
  \label{fig:bacc_vs_fms} 
\end{figure}

Table~\ref{tab:final_results_table} shows the results of the different machine learning methods on the two test sets. The results correspond to the representative of each method that achieved the best cross-validation accuracy, as described in Section~\ref{sec:experiments}. To account for the different frequencies of each primitive in the data, we report balanced accuracy, defined as the average of the classification accuracies for each primitive\footnote{The accuracy for each primitive is defined as $c_i/n_i$, $i=1,2,\ldots,5$, where $n_i$ is the number of windows associated with the $i$th primitive, and $c_i$ denotes how many were classified correctly. The balanced accuracy is the average of the accuracies corresponding to the five primitives.}. The results without taking into account primitive frequency are very similar, see Table~\ref{tab:acc_test_n_sev_pats}. Our main conclusion is that identification of functional primitives from IMU-sensor data via machine learning is possible: the deep learning methods achieved between 64\% and 70\% balanced accuracy on mildly-moderately impaired patients that do not appear in the training or validation data (Test set 1). On severely impaired patients (Test set 2) the average balanced accuracy was lower, between 38\% and 44\%. Decreased performance is expected because the training data only includes mildly and moderately impaired patients. Severely impaired patients not only had altered motion characteristics relative to less impaired patients, but their motions were also ascertained using modified activities adapted to their impairment level, as described in Section~\ref{sec:app_desc_act}. We used linear regression model to examine the effect of biological variables on model performance. We found no significant effects of gender, race, ethnicity, age, or impairment score on classification performance in the test set 1. Figure~\ref{fig:bacc_vs_fms} shows the results of the ensemble model on the individual patients in the two test sets. The balanced accuracy was above 60\% for all mildly-moderately impaired patients, and above 50\% for three out of the seven severely impaired patients (for comparison, random assignment yields 20\% accuracy).  

\begingroup
\renewcommand*{\arraystretch}{1.25}
\begin{table}[t]
\begin{center}
\begin{tabular}{|c|c|c|c|c|}
\hline
Architecture & \multicolumn{2}{c|}{ResNet} & \multicolumn{2}{c|}{DenseNet} \\ \hline
Normalization & BN & IN & BN & IN \\ \hline
Input embedding & 66.57  & 69.21  & 65.78 & 68.11  \\ \hline
No input embedding & 63.50 & 66.12 & 64.01  & 66.66 \\ \hline
\end{tabular}


\caption{Ablation analysis evaluating the individual contributions of input embeddings and instance normalization for two different convolutional architectures, described in more detail in Section~\ref{sec:app_schematic_conv}. The entries indicate the balanced accuracy of the different models on Test set 1 (mildly and moderately-impaired patients). BN denotes batch normalization, and IN denotes instance normalization. For both architectures the input embeddings and the adaptive normalization independently increased accuracy by 2-3\%. When combined, the increase was 4-5\%.}
\label{tab:conv_models_table}
\end{center}
\end{table}
\endgroup

Deep learning methods, which process the multivariate time series directly, systematically outperformed the techniques based on statistical features. Among the baseline deep learning methods, the recurrent network (LSTM) produced better results than the convolutional network. The best results overall were achieved by our proposed model, a convolutional network that incorporates input embeddings and instance normalization, as described in Section~\ref{sec:methodology}. An ensemble of this network and the LSTM, computed by averaging their outputs, produced a slight improvement. In Table~\ref{tab:conv_models_table} we show the results of an ablation analysis evaluating the individual contributions of input embeddings and instance normalization for two different convolutional architectures. For both architectures, the input embeddings and the adaptive normalization independently increased accuracy by 2-3\%. When combined, the increase was 4-5\%. The same trend was observed during validation on the different cross-validation folds, as shown in Table~\ref{tab:conv_val_perf}.   

\begin{figure}[t]
\begin{tabular}{>{\centering\arraybackslash}m{0.47\linewidth}  >{\centering\arraybackslash}m{0.47\linewidth}  } 
\includegraphics[width=1.05\linewidth]{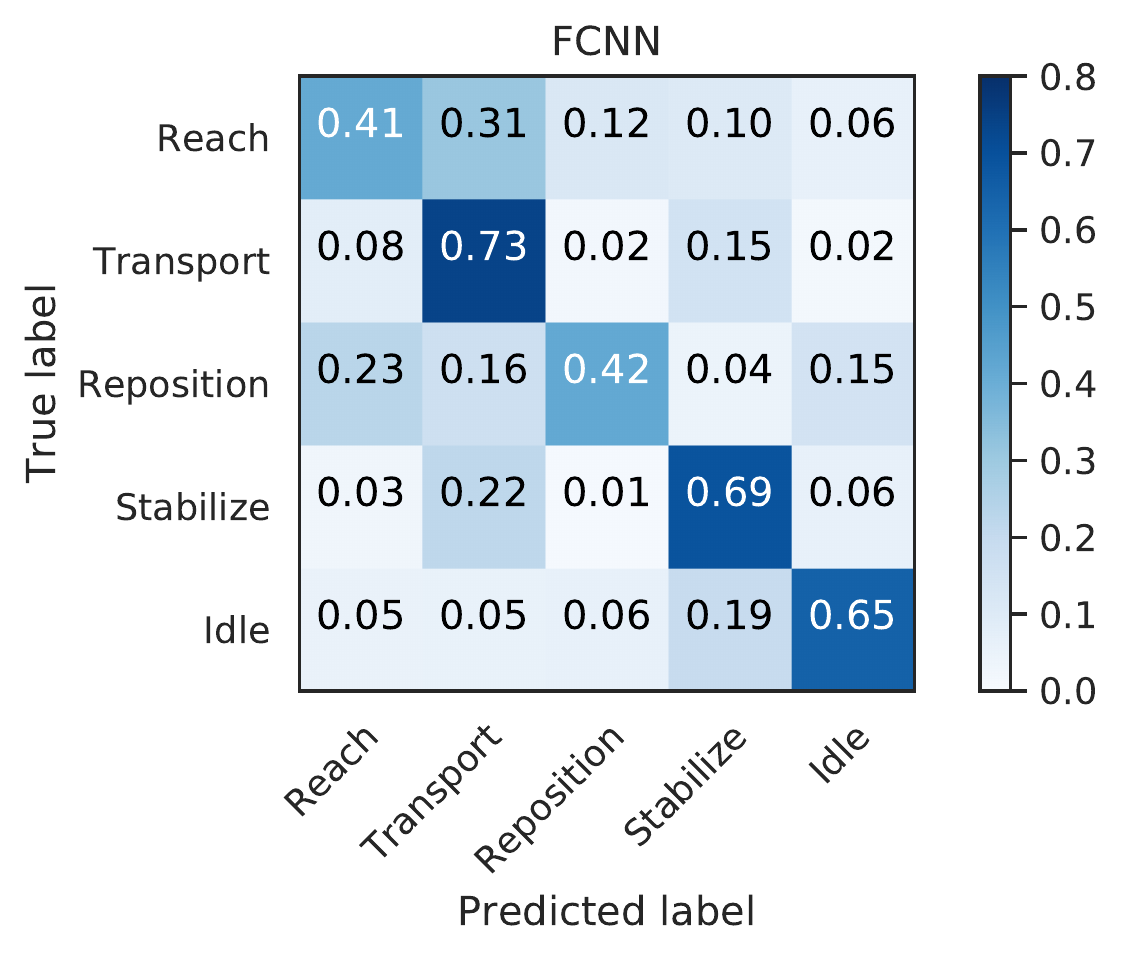}  & \includegraphics[width=1.05\linewidth]{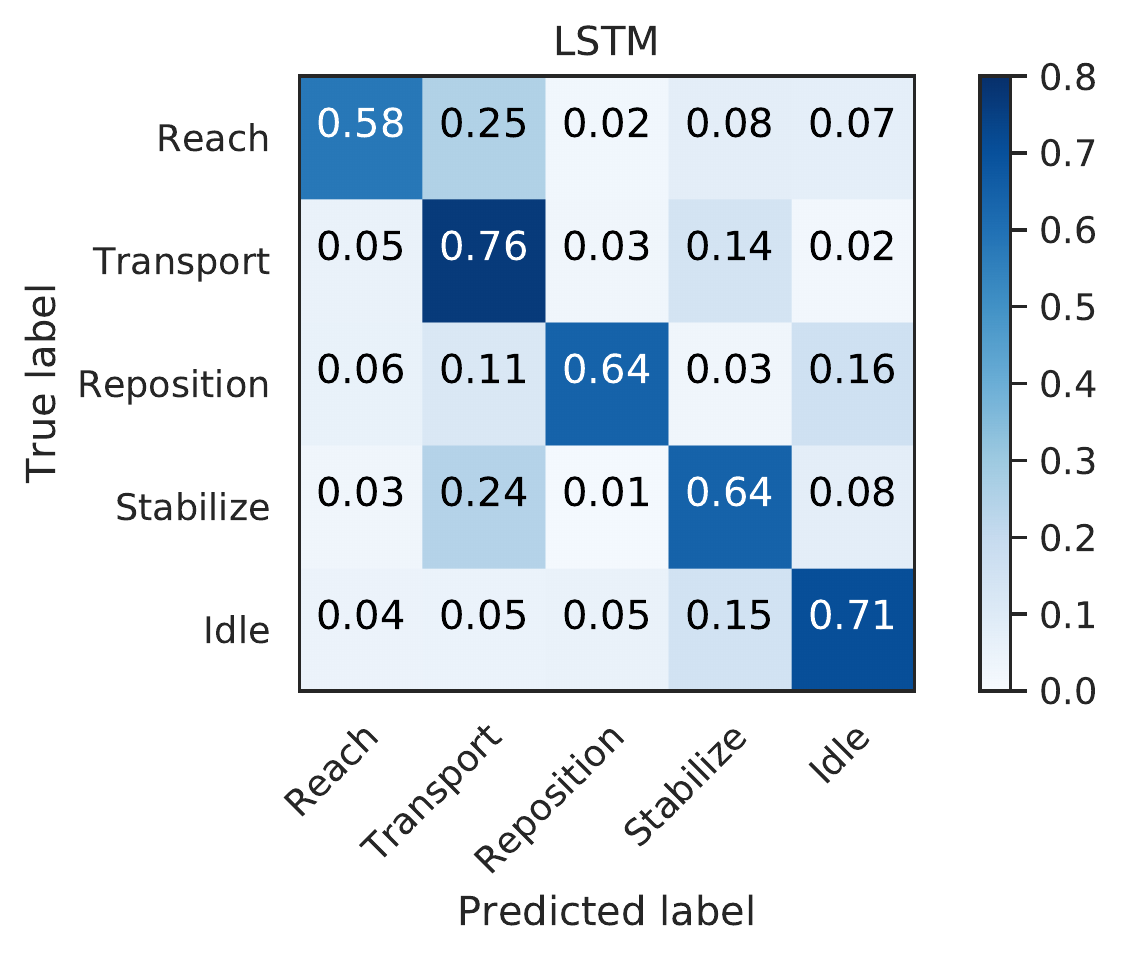} \\
\includegraphics[width=1.05\linewidth]{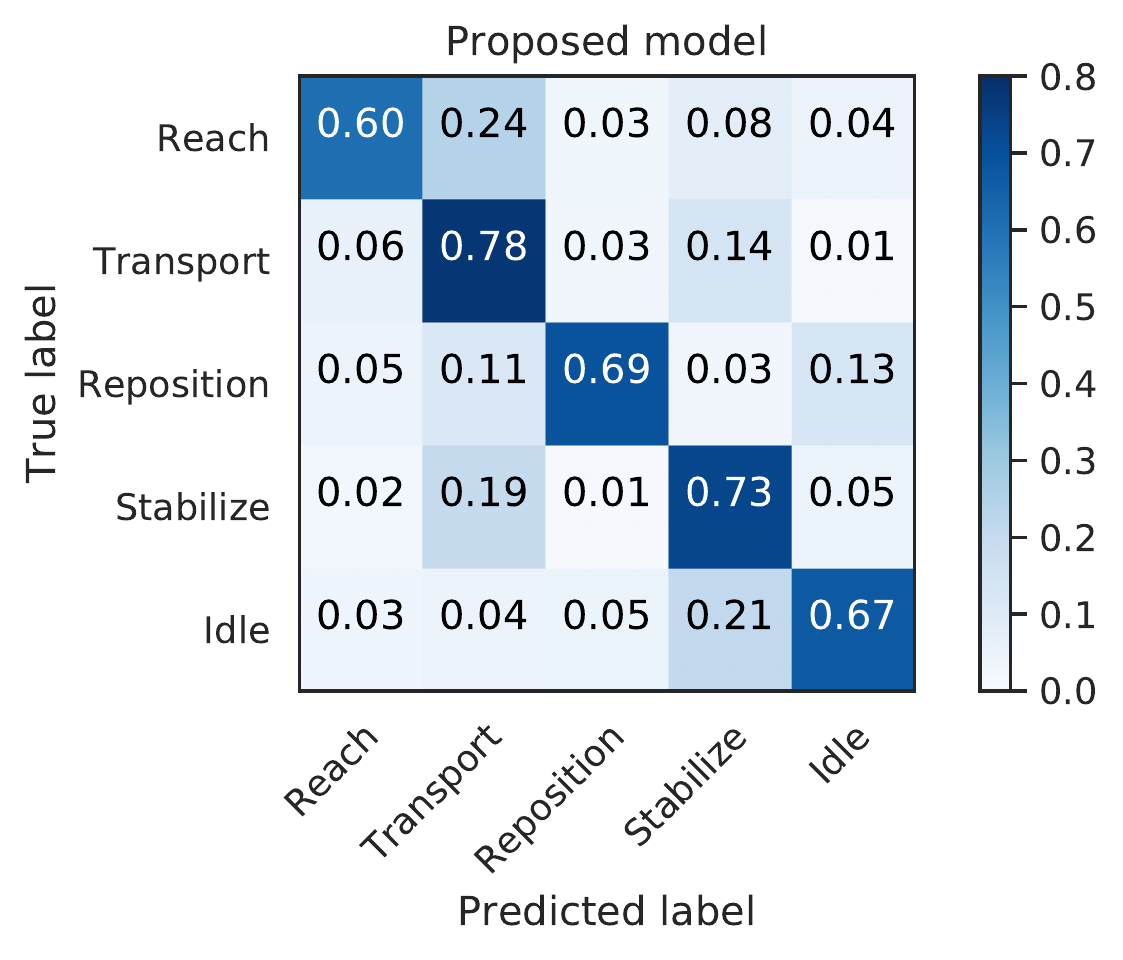}  & \includegraphics[width=1.05\linewidth]{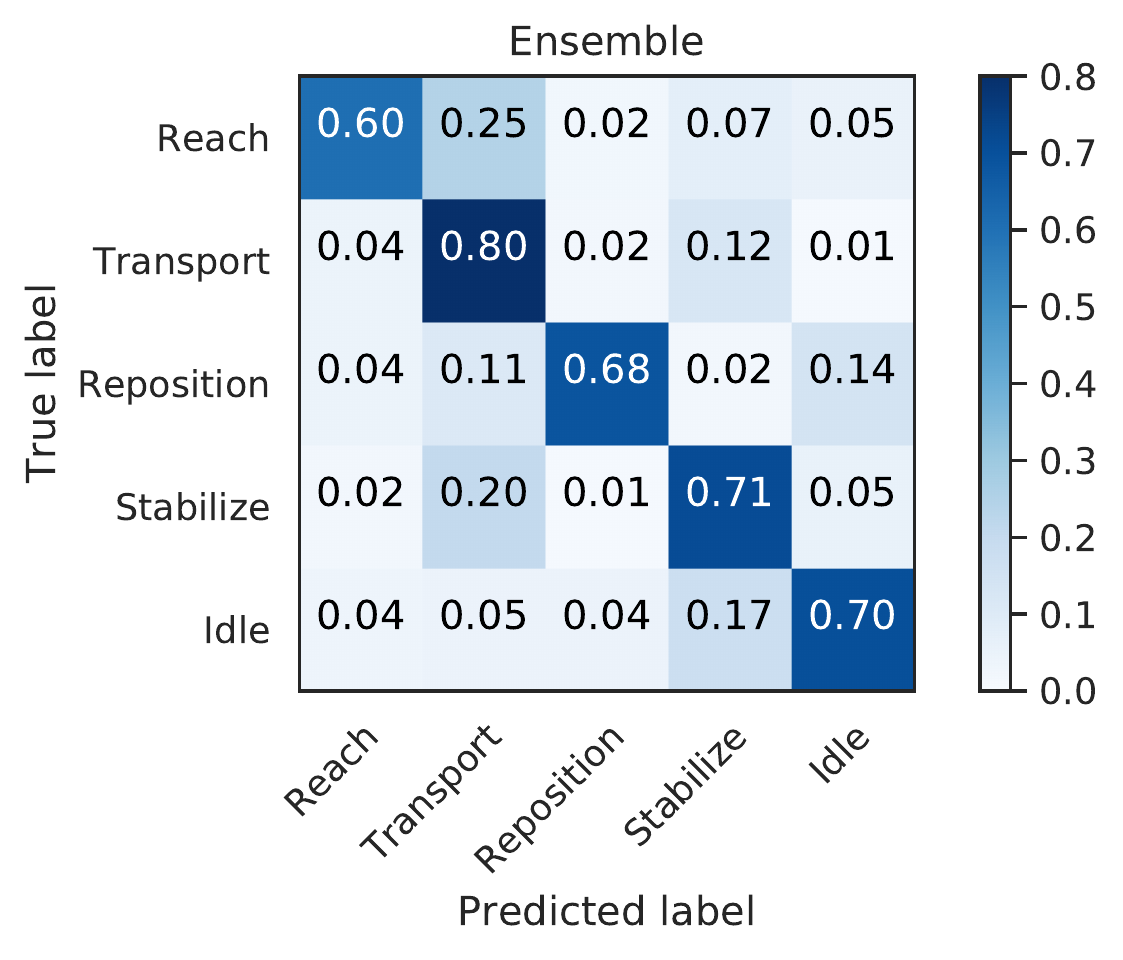}
\end{tabular}
\caption{Confusion matrices of the fully connected neural network (FCNN), the LSTM, the proposed model, and an ensemble of the proposed model and the LSTM on Test set 1 (mildly and moderately impaired patients). Each entry indicates what fraction of windows labeled with \emph{True label} were assigned to \emph{Predicted label} by each model. The models had similar error patterns, indicating that some primitive pairs are inherently more difficult to distinguish (e.g. reach and transport, idle and stabilize, stabilize and transport).
  }
  \label{fig:combined_cf_mat} 
\end{figure}


\begin{figure}[t]
  \centering 
  \includegraphics[width=\linewidth]{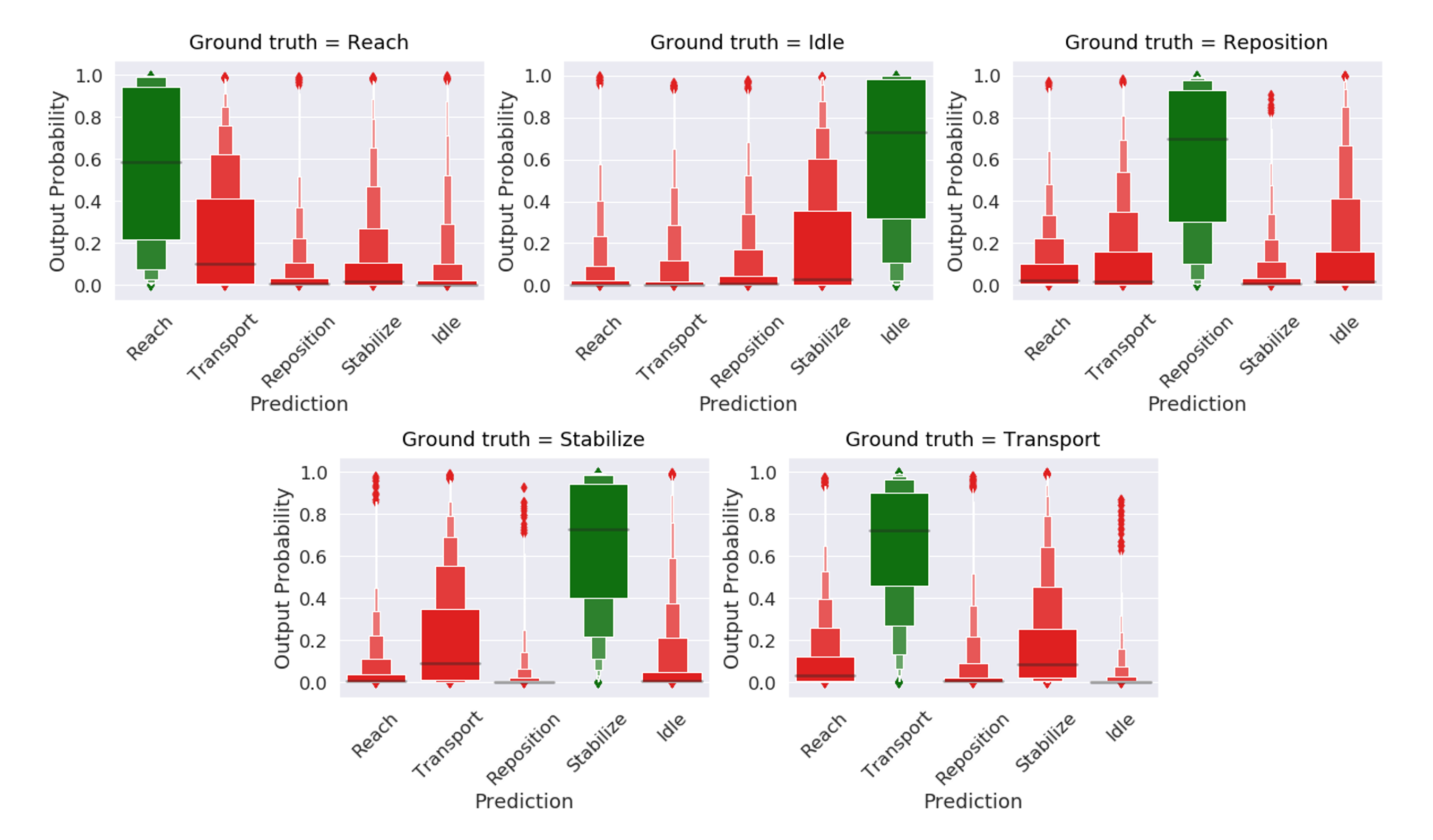} 
  \caption{Letter-value plots or Boxen plots~\citep{hofmann2017value} of the probability estimates generated by the ensemble model for the different primitives. The Boxen plots corresponding to the ground-truth primitive are colored in green, the rest are colored in red. At least half of the probabilities assigned to the ground-truth primitive are above 0.6. In contrast, the vast majority of the probabilities corresponding to other primitives are less than 0.6. This suggests that the probability estimate produced by the model is informative about its accuracy.}
  \label{fig:bp_combined_1} 
\end{figure}


Figure~\ref{fig:combined_cf_mat} shows the confusion matrices of several of the methods on Test set 1. The different models had similar error patterns, indicating that some primitive pairs are inherently more difficult to distinguish. Some of these errors, e.g. between reach and transport or idle and stabilize, may result from the lack of grasp information in the data. The proposed model had the highest accuracy for all primitives except idle. The ensemble model, combining the proposed model with the LSTM, improved accuracy on the idle and transport primitives, but also decreased it slightly for reach and reposition. Figure~\ref{fig:bp_combined_1} displays the probability estimates generated by the ensemble model applied to Test set 1 in the form of letter-value plots or Boxen plots~\citep{hofmann2017value}. These plots show the quantiles of the probabilities, the middle line corresponds to the median. At least half of the probabilities assigned to the correct primitive (green Boxen plots) are above 0.6. In contrast, the vast majority of the probabilities corresponding to other primitives (red Boxen plots) are less than 0.6. This suggests that the probability estimate produced by the model is informative about its accuracy.

\begin{figure}[t]
\hspace{-0.4cm}
\begin{tabular}{>{\centering\arraybackslash}m{0.32\linewidth}  >{\centering\arraybackslash}m{0.32\linewidth} >{\centering\arraybackslash}m{0.32\linewidth}} 
(a) & (b) & (c)\\
\includegraphics[width=\linewidth]{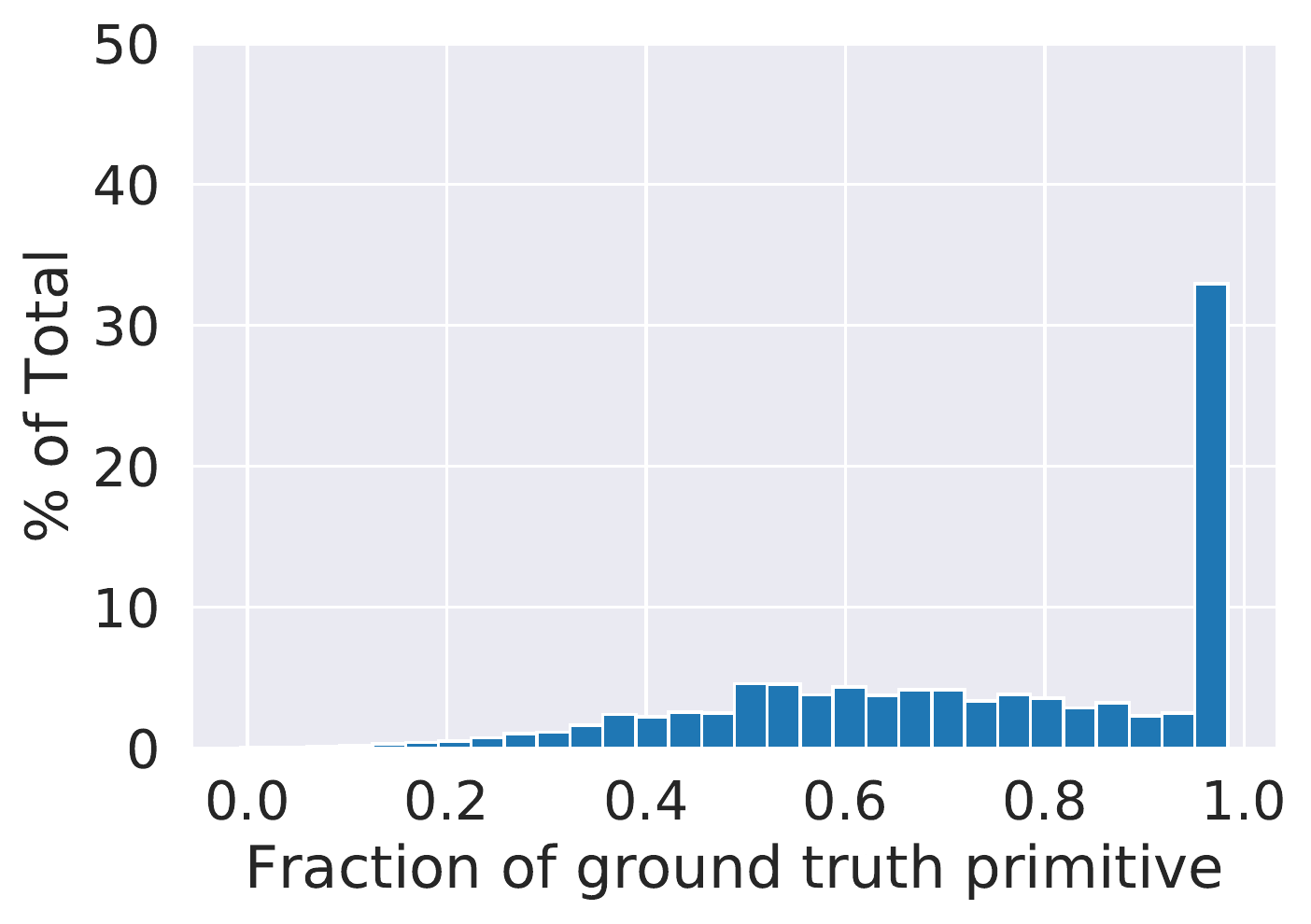}  & \includegraphics[width=\linewidth]{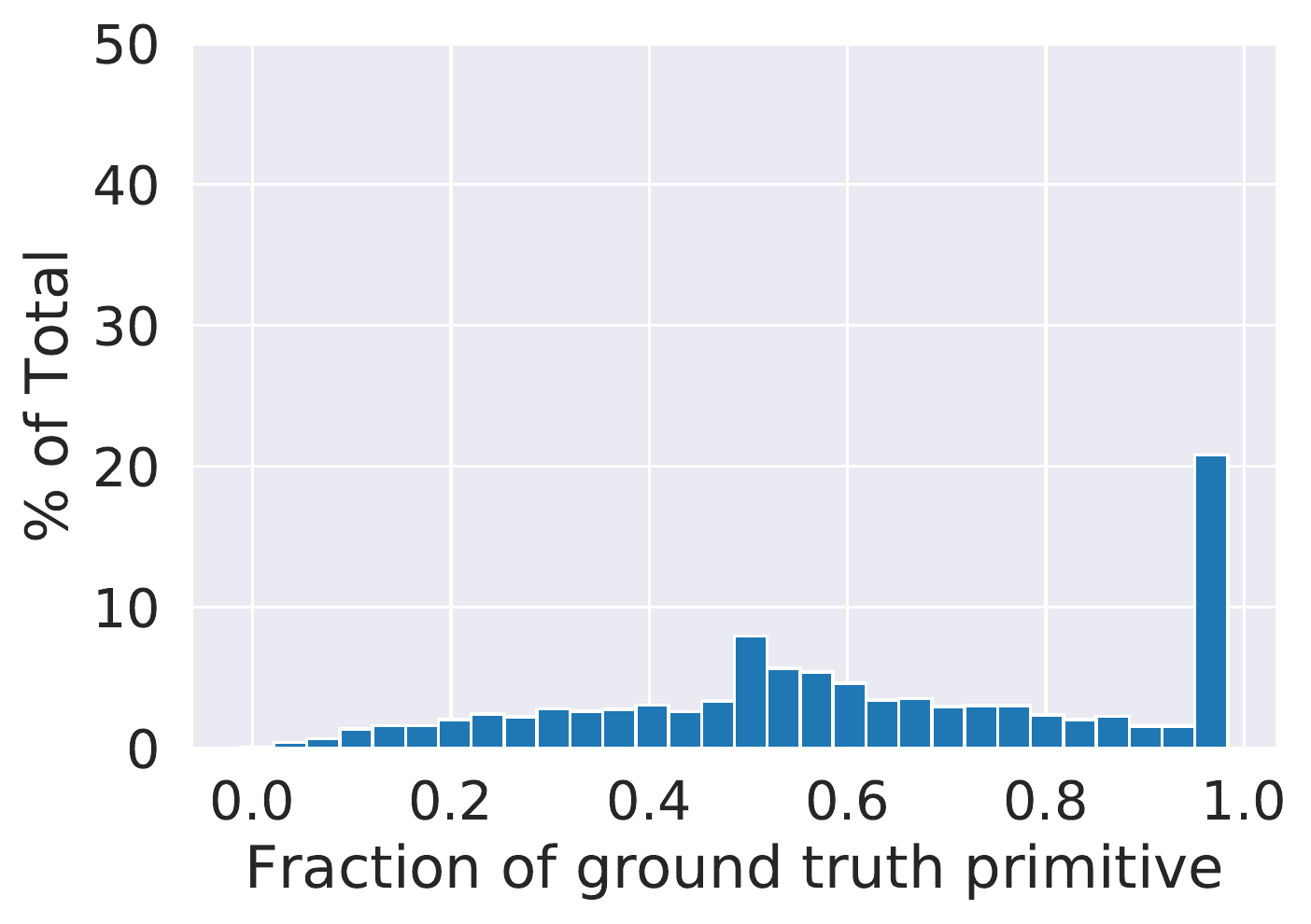} &
\includegraphics[width=\linewidth]{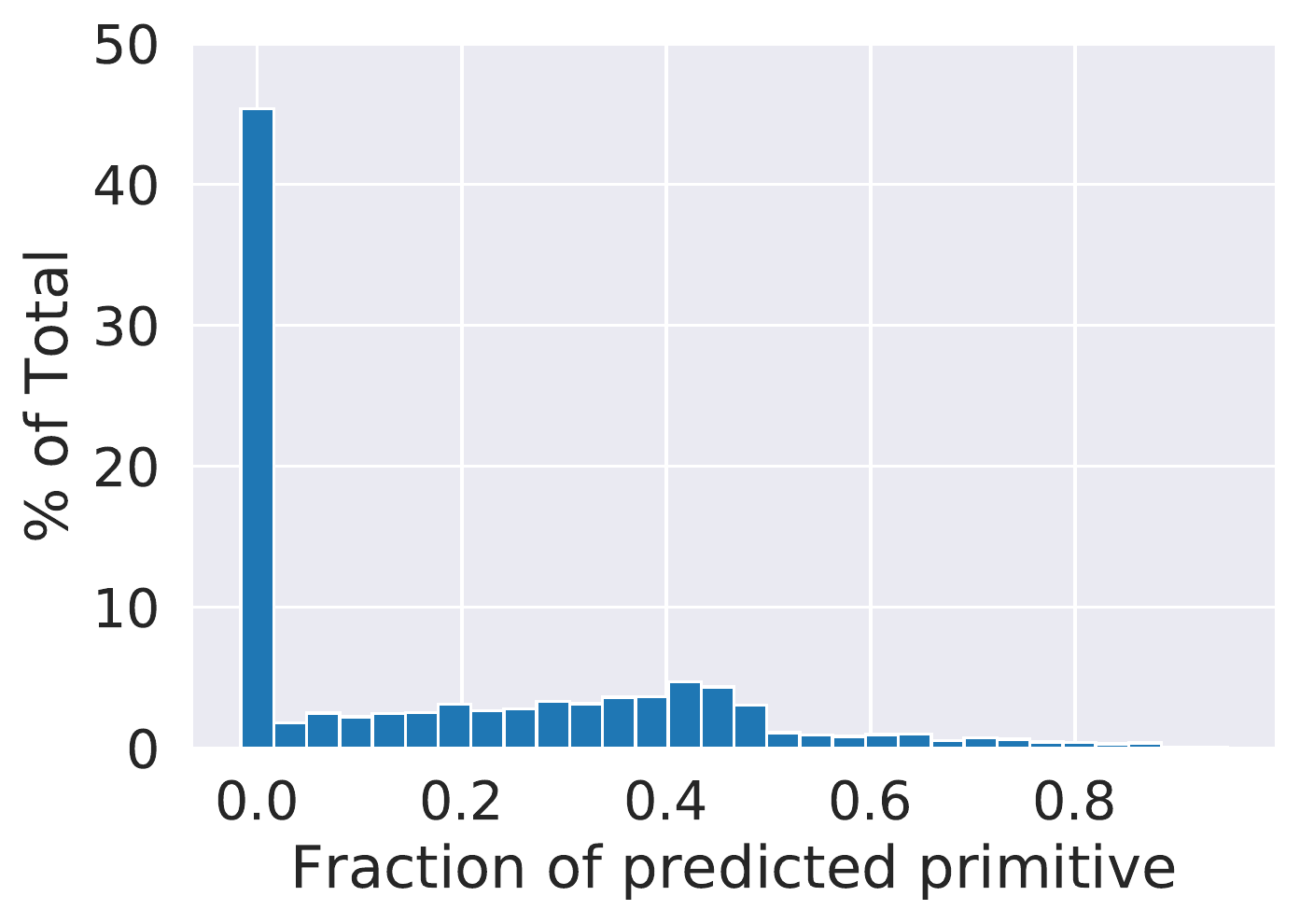} 
\end{tabular}
\caption{Histograms of the composition of the windows in Test set 1, examining the relationship between the classification results and the labels of the surrounding time steps. The ground-truth primitive of a window is defined as the label associated with the center of the window. The predicted primitive is the estimate produced by the ensemble model. (a) Percentage of time steps in each window associated with the ground-truth primitive for correctly-classified instances. (b) Percentage of time steps in the window associated with the ground-truth primitive for incorrectly-classified instances. (c) Percentage of time steps in each window associated with the predicted primitive for incorrectly-classified instances. Incorrectly-classified windows tend to contain a smaller fraction of the ground-truth primitive, but the difference between the histograms (compare a and b) is not very pronounced. This suggests that the ensemble model was relatively robust to the presence of additional primitives. In fact, more than two thirds of the windows that were correctly classified contained additional primitives. Among windows that were classified incorrectly, 45\% of them did not contain the predicted primitive (c).}
\label{fig:histogram}
\end{figure} 

Recall that we frame primitive identification as a classification problem, where the input is a 2-second window of sensor data and the label is the primitive associated with the center of the window, which we dub the \emph{ground-truth} primitive. A significant fraction of the window may contain different primitives, which is a potential source of errors. The machine learning models may be fooled by the other primitives and fail to detect the ground-truth primitive. Figure~\ref{fig:histogram} shows the composition of windows in Test set 1, separated depending on whether they were classified correctly or incorrectly. Incorrectly classified windows tend to contain a smaller fraction of the ground-truth primitive, but the difference between the histograms (compare a and b) is not very pronounced. This suggests that the ensemble model was relatively robust to the presence of additional primitives. In fact, more than two thirds of the windows that were correctly classified contained additional primitives.  

\section{Discussion} \label{sec:discussion}
This study demonstrates that deep learning can be used to identify functional primitives from IMU sensor data, which is an important step towards developing quantitative approaches for measuring stroke rehabilitation. It also suggests that input embeddings and adaptive feature normalization may contribute to address two challenges arising in many healthcare applications of machine learning: processing data containing different physical quantities, and ensuring robustness to distributional shifts during inference. 

The classification performance of our approach exceeds random chance (20\%) and is comparable to the accuracy (70\%) of an approach that dichotomizes motion into time spent in functional versus nonfunctional motion \citep{bochniewicz2017measuring}. Importantly, our approach identifies the content of functional motion, i.e. functional primitives, that will serve as the basis for detailed rehabilitation measurement. Still, we anticipate that additional gains in classification performance can be made. We observed that the models had difficulty distinguishing reaches from transports, and idles from stabilizations. These primitives differ by the presence and timing of grasp \citep{schambra2019taxonomy}, indicating that our current IMU array does not communicate this level of detail. However, affixing additional IMUs to paretic fingers would hinder hand function and further limit the practical utility of the approach. A recently developed computer vision model may offer a solution: it can extract finger position from video recordings \citep{cao2018openpose}. This new capability could enable us to use our existing video dataset to retrieve information about grasp. Future work will test whether combining kinematic information from IMUs and cameras can effectively boost classification accuracy. 

Our study has some limitations to be considered. We studied only right-dominant patients balanced for right and left paresis. This step was necessary to simplify classification. Hand dominance may have a differential influence on the preferential roles of the UEs and their kinematic signatures \citep{przybyla2012dynamic}. As the majority of humans are right-dominant, the proposed approach would be applicable to most patients. In the future, the inclusion of left-dominant patients for training and testing would enable us to build a more universal tool. 

Another limitation of our approach is that classification performance deteriorates significantly for severely impaired patients, which means that it cannot be safely generalized to this cohort. This observation opens up two avenues for future work. First, gathering larger datasets, we will be able to include more severely impaired patients to train our models. Second, we will aim to develop machine-learning methodology capable of generalizing more robustly to different levels of impairment.

Finally, we performed primitive identification at a high time granularity (time steps of 10 ms). Future work will focus on converting these predictions to a sequence of estimated primitives. This may be expected to enable the next step in our approach, which is to automatically count primitives after they have been successfully recognized. 

In summary, we present an approach that combines the kinematic data from IMUs with optimized deep learning models to identify functional primitives that constitute rehabilitation activities. We envision that once classification performance is maximized for an array of impairment, the trained model can be deployed to a clinical setting. There, patients instrumented with IMUs will undergo rehabilitation, generating unlabeled kinematic data. Using these data, the trained model will extract primitive content and count. This approach is expected to provide an objective means of quantitating the training dose of stroke rehabilitation. This measurement ability opens up a path for critical dose-response research and informed delivery of dosed rehabilitation, vital for improving recovery outcomes in stroke patients.



\acks{
We would also like to thank the volunteers who contributed to label the dataset: Ronak Trivedi, Adisa Velovic, Sanya Rastogi, Candace Cameron, Sirajul Islam, Bria Bartsch, Courtney Nilson, Vivian Zhang, Nicole Rezak, Christopher Yoon, Sindhu Avuthu, and Tiffany Rivera. We thank Dawn Nilsen, OT EdD for expert advice on the testing battery, and Audre Wirtanen for early assistance with the testing setup and data collection. This work was supported by an AHA postdoctoral fellowship 19AMTG35210398 (AP), NIH grants R01 LM013316 (AK, CFG, HMS) and K02 NS104207 (HMS), NSF NRT-HDR Award 1922658 (CFG) and by the Moore-Sloan Data Science Environment at NYU (AK).}

\bibliography{mlhc_sample}

\appendix

\section{Description of the Rehabilitation Activities}  
\label{sec:app_desc_act}
Tables~\ref{tab:act_descp_1} and \ref{tab:act_descp_2} describe the activities performed by the mildly and moderately impaired stroke patients in the cohort. Tables~\ref{tab:sev_act_descp_1} and \ref{tab:sev_act_descp_2} describe the activities performed by the severely impaired patients assigned to Test set 2.
\begingroup
\renewcommand*{\arraystretch}{1}
\newcolumntype{C}[1]{>{\centering\arraybackslash}p{#1}}
\begin{table}[!htbp]
\resizebox{\textwidth}{!}{%
\centering
\begin{tabular}{|p{2cm}|p{5cm}|p{2cm}|p{5cm}|}
\hline
\multicolumn{1}{|c|}{Activity} & \multicolumn{1}{c|}{Workspace} & \multicolumn{1}{c|}{\begin{tabular}[c]{@{}c@{}}Target\\ object(s)\end{tabular}} & \multicolumn{1}{c|}{Instructions} \\ 
\hline
Washing face & Sink with a small tub in it and two folded washcloths on either side of the countertop, 30 cm from edge closest to patient & Washcloths, faucet handle & Fill tub with water, dip washcloth on the right side into water, wring it, wiping each side of their face with wet washcloth, place it back on countertop. Use washcloth on the left side to dry face, place it back on countertop \\ \hline
Applying deodorant & Tabletop with deodorant placed at midline, 25 cm from edge closest to patient & Deodorant & Remove cap, twist base a few times, apply deodorant, replace cap, untwist the base, put deodorant on table \\ \hline
Hair combing & Tabletop with comb placed at midline, 25 cm from edge closest to patient & Comb & Pick up comb and comb both sides of head \\ \hline
Don/doffing glasses & Tabletop with glasses placed at midline, 25 cm from edge closest to patient & Glasses & Wear glasses, return hands to table, remove glasses and place on table\\ \hline
Eating & Table top with a standard-size paper plate (at midline, 2 cm from edge), utensils (3 cm from edge, 5 cm from either side of plate), a baggie with a slice of bread (25 cm from edge, 23 cm left of midline), and a margarine packet (32 cm from edge, 17 cm right of midline) & Fork, knife, re-sealable sandwich baggie, slice of bread, single-serve margarine container & Remove bread from plastic bag and put it on plate, open margarine pack and spread it on bread, cut bread into four pieces, cut off and eat a small bite-sized piece \\ \hline
\end{tabular}}
\caption{Description of the activities performed by the mildly and moderately impaired patients in the cohort (1/2).}
\label{tab:act_descp_1}
\end{table}

\begin{table}[!htbp]
\resizebox{\textwidth}{!}{%
\centering
\begin{tabular}{|p{2cm}|p{5cm}|p{2cm}|p{5cm}|}
\hline
\multicolumn{1}{|c|}{Activity} & \multicolumn{1}{c|}{Workspace} & \multicolumn{1}{c|}{\begin{tabular}[c]{@{}c@{}}Target\\ object(s)\end{tabular}} & \multicolumn{1}{c|}{Instructions} \\ \hline
Drinking & Tabletop with water bottle and paper cup 18 cm to the left and right of midline, 25 cm from edge closest to patient & Water bottle (12 oz),
paper cup & Open water bottle, pour water into cup, take a sip of water, place cup on table, and replace cap on bottle \\ \hline
Tooth brushing & Sink with toothpaste and toothbrush on either side of the countertop, 30 cm from edge closest to patient & Travel-sized toothpaste, toothbrush with built-up foam grip, faucet handle & Wet toothbrush, apply toothpaste to toothbrush, replace cap on toothpaste tube, brush teeth, rinse toothbrush and mouth, place toothbrush back on countertop \\ \hline
Moving object on a horizontal surface & Horizontal circular array (48.5 cm diameter) of 8  targets (5 cm diameter) & Toilet paper roll & Move the roll between the center and each outer target, resting between each motion and at the end \\ \hline
Moving object on/off a Shelf & Shelf with two levels (33 cm and 53 cm) with 3 targets on both levels (22.5 cm, 45 cm, and 67.5 cm away from the left-most edge) & Toilet paper roll & Move the roll between the center target and each target on the shelf, resting between each motion and at the end \\ \hline
\end{tabular}}
\caption{Description of the activities performed by the mildly and moderately impaired patients in the cohort (2/2).}
\label{tab:act_descp_2}
\end{table}

\endgroup

\begingroup
\renewcommand*{\arraystretch}{1}
\begin{table}[!htbp]
\resizebox{\textwidth}{!}{%
\begin{tabular}{|p{2cm}|p{3.7cm}|p{2cm}|p{3.35cm}|p{3.35cm}|}
\hline
\multicolumn{1}{|c|}{\multirow{2}{*}{Activity}} & \multicolumn{1}{c|}{\multirow{2}{*}{Workspace}} & \multicolumn{1}{c|}{\multirow{2}{*}{\begin{tabular}[c]{@{}c@{}}Target \\ object(s)\end{tabular}}} & \multicolumn{2}{c|}{Instructions $^*$} \\ \cline{4-5} 
\multicolumn{1}{|c|}{} & \multicolumn{1}{c|}{} & \multicolumn{1}{c|}{} & \multicolumn{1}{c|}{Proximal \textgreater Distal} & \multicolumn{1}{c|}{Proximal \textless Distal} \\ \hline
 Washing face & Sink with a small tub in it and two folded washcloths on either side of the countertop, 30 cm from edge closest to patient & Washcloths, faucet handle & Reach to touch faucet knob. Place washcloth in paretic hand and bring to both sides of face. & Open and close faucet. Lift washcloth from basin and wring it out.  \\ \hline
 Applying deodorant & Tabletop with deodorant placed at midline, 25 cm from edge closest to patient & Deodorant & Reach to touch deodorant. Place deodorant in paretic hand and bring to opposite armpit. & Lift deodorant for 3 seconds. From the horizontal position, rotate deodorant upright and return to original position.  \\ \hline
 Hair combing & Tabletop with comb placed at midline, 25 cm from edge closest to patient & Comb & Reach to touch comb. Place comb in paretic hand and bring to both sides of head. & Lift comb for 3 seconds. \\ \hline
 Don/doffing glasses & Tabletop with glasses placed at midline, 25 cm from edge closest to patient & Glasses & Reach to touch glasses. & Lift glasses for 3 seconds. \\ \hline
Eating & Table top with a standard-size paper plate (at midline, 2 cm from edge), utensils (3 cm from edge, 5 cm from either side of plate), a baggie with a slice of bread (25 cm from edge, 23 cm left of midline), and a margarine packet (32 cm from edge, 17 cm right of midline) & Fork, knife, re-sealable sandwich baggie, slice of bread, single-serve margarine container & Reach to touch each item separately on paretic side. Place fork in paretic hand and bring fork to mouth. & Lift each object on paretic side for 3 seconds. \\ \hline
\end{tabular}}
\caption{Description of the activities performed by the severely impaired patients in the cohort (1/2). $^*$ Instructions for the severely impaired patients were given based on the UE segment with greater preserved function. “Proximal $>$ distal” indicates better strength in the proximal (i.e. deltoid, biceps, triceps) than distal (i.e. hand) UE, which was typically paralyzed in these patients. The initial UE position was generally at the edge of the table/counter closest to the patient. “Distal $>$ proximal” had the opposite distribution of strength. The initial UE position was adjacent to the target object. All testing were done on the paretic UE.}
\label{tab:sev_act_descp_1}
\end{table}

\begin{table}[!htbp]
\small
\resizebox{\textwidth}{!}{%
\begin{tabular}{|p{2cm}|p{3.7cm}|p{2cm}|p{3.4cm}|p{3.4cm}|}
\hline
\multicolumn{1}{|c|}{\multirow{2}{*}{Activity}} & \multicolumn{1}{c|}{\multirow{2}{*}{Workspace}} & \multicolumn{1}{c|}{\multirow{2}{*}{\begin{tabular}[c]{@{}c@{}}Target \\ object(s)\end{tabular}}} & \multicolumn{2}{c|}{Instructions$^*$} \\ \cline{4-5} 
\multicolumn{1}{|c|}{} & \multicolumn{1}{c|}{} & \multicolumn{1}{c|}{} & \multicolumn{1}{c|}{Proximal \textgreater Distal} & \multicolumn{1}{c|}{Proximal \textless Distal} \\ \hline
Drinking & Tabletop with water bottle and paper cup 18 cm to the left and right of midline, 25 cm from edge closest to patient & Water bottle (12 oz), paper cup & Reach to touch object on paretic side. Reach across to touch object on non-paretic side. & Starting from upright position, lay object on paretic side horizontally, release, and return to upright. Perform same series of actions on the object on the non-paretic side. \\ \hline
Tooth brushing&Sink with toothpaste and toothbrush on either side of the countertop, 30 cm from edge closest to patient&Travel-sized toothpaste on left, toothbrush with built-up foam grip on right, faucet handle&Reach to touch object on paretic side. Place toothbrush in paretic hand and bring it to mouth.&Lift object on paretic side for 3 seconds. \\ \hline
Moving object on a horizontal surface&Horizontal circular array (48.5 cm diameter) of 8  targets (5 cm diameter)&Toilet paper roll (200 g) or can (200 g)&Investigator will assess if toilet paper roll can be grasped, or aluminum can if not. 
If grasp is possible, move roll/can between the center and each outer target, resting before and after each motion. 

If grasp is not possible, the toilet paper roll will be moved around the target array by the investigator and the patient will reach to touch it at each location. 
& Investigator will assess if toilet paper roll can be grasped, or aluminum can if not. 

If grasp is not possible, the toilet paper roll will be moved around the target array by the investigator and the patient will reach to touch it at each location.\\ \hline
Moving object on/off a Shelf& Shelf with two levels (33 cm and 53 cm) with 3 targets on both levels (22.5 cm, 45 cm, and 67.5 cm away from the left-most edge)&Toilet paper roll (200 g) or can (200 g)&Investigator will assess if toilet paper roll can be grasped, or aluminum can if not. 
If grasp is possible, move roll/can between the center and each outer target, resting before and after each motion. 

If grasp is not possible, the toilet paper roll will be moved around the target array by the investigator and the patient will reach to touch it at each location. 
&Investigator will assess if toilet paper roll can be grasped, or aluminum can if not. 

If grasp is not possible, the toilet paper roll will be moved around the target array by the investigator and the patient will reach to touch it at each location. \\ \hline
\end{tabular}}
\caption{Description of the activities performed by the severely impaired patients in the cohort (2/2).}
\label{tab:sev_act_descp_2}
\end{table}
\endgroup

\section{Description of the Joint Angles}  \label{sec:app_joint_angles}
As described in Section~\ref{sec:data}, the sensor measurements are used to compute 22 anatomical angle values using a rigid-body skeletal model scaled to the patient's height and segment lengths. Table \ref{tab:joint_angles} describes these joint angles in  detail.
\begingroup
\renewcommand*{\arraystretch}{1}
\begin{table}[!htbp]
\centering
\begin{tabular}{|l|l|}
\hline
Joint/segment & Anatomical angle \\ \hline
Shoulder & \begin{tabular}[c]{@{}l@{}}Shoulder flexion/extension\\ Shoulder internal/external rotation\\ Shoulder ad-/abduction\\ Shoulder total flexion$^\ddagger$ \end{tabular} \\ \hline
Elbow & Elbow flexion/extension \\ \hline
Wrist & \begin{tabular}[c]{@{}l@{}}Wrist flexion/extension\\ Forearm pronation/supination\\ Wrist radial/ulnar deviation\end{tabular} \\ \hline
Thorax & \begin{tabular}[c]{@{}l@{}}Thoracic$^*$ flexion/extension\\ Thoracic$^*$ axial rotation\\ Thoracic$^*$ lateral flexion/extension\end{tabular} \\ \hline
Lumbar & \begin{tabular}[c]{@{}l@{}}Lumbar$^\dagger$ flexion/extension\\ Lumbar$^\dagger$ axial rotation\\ Lumbar$^\dagger$ lateral flexion/extension\end{tabular} \\ \hline
\end{tabular}
\caption{List of anatomical angles. The system uses a rigid-body skeletal model to convert the IMU measurements into joint and segment angles. $\ddagger$ Shoulder total flexion is a combination of shoulder flexion/extension and shoulder ad-/abduction. $^*$Thoracic angles are computed between the cervical vertebra and the thoracic vertebra. $\dagger$Lumbar angles are computed between the thoracic vertebra and pelvis. }
\label{tab:joint_angles}
\end{table}
\endgroup

\section{Additional Results} \label{sec:additional_results}
Table \ref{tab:acc_test_n_sev_pats} shows the performance of the different machine-learning models on Test sets 1 and 2 measured using accuracy instead of balanced accuracy. We also provide the cross-validation results on the individual splits, and the average validation accuracy, in Tables~\ref{tab:fcnn_val_perf_dp_none}, \ref{tab:fcnn_val_perf_dp_02} and \ref{tab:fcnn_val_perf_dp_05} for the fully-connected neural network models, in Table~\ref{tab:lstm_val_perf} for the LSTM, and in Table~\ref{tab:conv_val_perf} for the convolutional models.
\begingroup
\renewcommand*{\arraystretch}{1}
\begin{table}[!htbp]
\begin{center}
\begin{tabular}{|c|c|c|c|c|c|c|}
\hline
\begin{tabular}[c]{@{}c@{}}Method\end{tabular} & \begin{tabular}[c]{@{}c@{}}Random forest\end{tabular} & FCNN & \begin{tabular}[c]{@{}c@{}} CNN\end{tabular} & LSTM & \begin{tabular}[c]{@{}c@{}}Proposed \end{tabular} & Ensemble \\ \hline
\begin{tabular}[c]{@{}c@{}}Test set 1\end{tabular} & 59.66 & 62.43 & 64.98 & 68.21 & 70.67 & 71.87 \\ \hline
\begin{tabular}[c]{@{}c@{}}Test set 2\end{tabular} & 33.79 & 39.62 & 43.11 & 48.78 & 44.44 & 48.36 \\ \hline
\end{tabular}
\end{center}
\caption{Accuracy on Test set 1 and Test set 2 of the machine-learning models described in Section \ref{sec:experiments}. FCNN denotes fully connected neural network. The ensemble is a combination of the proposed model and the LSTM, where the output probabilities were averaged.}
\label{tab:acc_test_n_sev_pats}
\end{table}
\endgroup
\begingroup
\renewcommand*{\arraystretch}{1}
\begin{table}[!htbp]
\begin{center}
\begin{tabular}{|c|c|c|c|c|c|c|}
\hline
L & S & Fold 1 & Fold 2 & Fold 3 & Fold 4 & Average \\ \hline
1 & 2 &  56.23 &  54.77 & 57.97 &  57.21 &  \textbf{56.55} \\ \hline
1 & 3000 &  56.31 &  54.73 &  58.05 &  57.06 &  56.54 \\ \hline
1 & 1500 &  56.27 &  54.80 &  58.08 &  56.99 &  56.53 \\ \hline
1 & 120  &  56.22 &  54.58 &  58.07 &  57.06 &  56.48 \\ \hline
1 & 700  &  56.29 &  54.72 &  57.92 &  56.95 &  56.47 \\ \hline
1 & 300  &  56.36 &  54.58 &  57.86 &  56.92 &  56.43 \\ \hline
1 & 50   &  56.19 &  54.56 &  57.92 &  56.82 &  56.37 \\ \hline
5 & 120  &  56.27 &  54.57 &  57.80 &  56.82 &  56.37 \\ \hline
5 & 300  &  56.23 &  54.62 &  57.73 &  56.80 &  56.34 \\ \hline
1 & 20   &  56.06 &  54.48 &  57.93 &  56.88 &  56.34 \\ \hline
\end{tabular}
\end{center}
\caption{Validation accuracies for the random-forest models. We report the top 10 performing models. L denotes the minimum number of samples required to be at a leaf and S denotes the minimum number of examples required to split each internal node. We experimented with the following values: L = \{1,5,20,50,150,400,800,1500\}, S = \{2,20,50,120,300,700,1500,3000\}.}
\label{tab:rf_val_perf}
\end{table}
\endgroup

\begingroup
\renewcommand*{\arraystretch}{1}
\begin{table}[!htbp]
\begin{center}
\begin{tabular}{|c|c|c|c|c|c|c|}
\hline
\# of layers & Dim. of hidden units & Fold 1 & Fold 2 & Fold 3 & Fold 4 & Average \\ \hline
4 & 300 & 56.86 & 55.47 & 56.85 & 55.49 & 56.17 \\ \hline
4 & 600 & 55.44 & 56.73 & 58.23 & 55.79 & 56.55 \\ \hline
4 & 900 & 56.02 & 56.89 & 58.13 & 54.83 & 56.47 \\ \hline
8 & 300 & 56.52 & 56.22 & 58.28 & 54.4 & 56.36 \\ \hline
\textbf{8} & \textbf{600} & \textbf{56.54} & \textbf{56.43} & \textbf{58} & \textbf{56.01} & \textbf{56.75} \\ \hline
8 & 900 & 57.05 & 56.27 & 57.31 & 55.68 & 56.58 \\ \hline
12 & 300 & 55.27 & 56.16 & 55.83 & 55.45 & 55.68 \\ \hline
12 & 600 & 56.23 & 56.51 & 57.68 & 54.47 & 56.22 \\ \hline
12 & 900 & 55.9 & 56.68 & 58 & 55.57 & 56.54 \\ \hline
\end{tabular}
\end{center}
\caption{Validation accuracy for fully-connected neural network models with different number of layers, different dimensions of hidden units, and no dropout.}
\label{tab:fcnn_val_perf_dp_none}
\end{table}
\endgroup

\begingroup
\renewcommand*{\arraystretch}{1}
\begin{table}[!htbp]
\begin{center}
\begin{tabular}{|c|c|c|c|c|c|c|}
\hline
\# of layers & Dim. of hidden units & Fold 1 & Fold 2 & Fold 3 & Fold 4 & Average \\ \hline
4 & 300 & 58.01 & 58.12 & 58.83 & 56.57 & 57.88 \\ \hline
4 & 600 & 56.78 & 58.06 & 59.07 & 55.80 & 57.43 \\ \hline
4 & 900 & 57.70 & 58.21 & 58.37 & 56.64 & 57.73 \\ \hline
8 & 300 & 57.70 & 57.64 & 58.47 & 56.32 & 57.53 \\ \hline
\textbf{8} & \textbf{600} & \textbf{57.41} & \textbf{58.62} & \textbf{59.22} & \textbf{56.67} & \textbf{57.98} \\ \hline
8 & 900 & 57.06 & 58.22 & 59.14 & 56.34 & 57.69 \\ \hline
12 & 300 & 57.43 & 57.49 & 58.78 & 55.70 & 57.35 \\ \hline
12 & 600 & 57.29 & 57.01 & 59.67 & 56.60 & 57.64 \\ \hline
12 & 900 & 57.40 & 57.97 & 58.90 & 55.68 & 57.49 \\ \hline
\end{tabular}
\end{center}
\caption{Validation accuracy for fully-connected neural network models with different number of layers, and different dimensions of hidden units. The dropout rate was set to 0.2 for all the models.}
\label{tab:fcnn_val_perf_dp_02}
\end{table}
\endgroup

\begingroup
\renewcommand*{\arraystretch}{1}
\begin{table}[!htbp]
\begin{center}
\begin{tabular}{|c|c|c|c|c|c|c|}
\hline
\# of layers & Dim. of hidden units & Fold 1 & Fold 2 & Fold 3 & Fold 4 & Average \\ \hline
4 & 300 & 58.20 & 58.19 & 59.62 & 57.16 & 58.29 \\ \hline
4 & 600 & 59.39 & 58.86 & 55.96 & 56.34 & 57.64 \\ \hline
4 & 900 & 58.43 & 58.57 & 60.34 & 57.25 & 58.65 \\ \hline
8 & 300 & 57.08 & 56.60 & 58.31 & 56.28 & 57.07 \\ \hline
8 & 600 & 58.14 & 58.59 & 60.08 & 57.22 & 58.51 \\ \hline
\textbf{8} & \textbf{900} & \textbf{58.08} & \textbf{58.86} & \textbf{60.33} & \textbf{57.38} & \textbf{58.66} \\ \hline
12 & 300 & 57.11 & 57.10 & 57.95 & 55.63 & 56.95 \\ \hline
12 & 600 & 57.81 & 58.14 & 58.91 & 56.24 & 57.78 \\ \hline
12 & 900 & 58.09 & 57.92 & 59.85 & 57.06 & 58.23 \\ \hline
\end{tabular}
\end{center}
\caption{Validation accuracy for fully-connected neural network models with different number of layers, and different dimensions of hidden units. The dropout rate was set to 0.5 for all the models.}
\label{tab:fcnn_val_perf_dp_05}
\end{table}
\endgroup

\begingroup
\renewcommand*{\arraystretch}{1}
\begin{table}[!htbp]
\begin{center}
\begin{tabular}{|c|c|c|c|c|c|}
\hline
Dim. of hidden units & Fold 1 & Fold 2 & Fold 3 & Fold 4 & Average \\ \hline
400 & 57.4 & 61.78 & 61.68 & 59.43 & 60.07 \\ \hline
1200 & 59.05 & 61.77 & 62.32 & 61.64 & 61.20 \\ \hline
2000 & 61.57 & 62.85 & 64.38 & 60.95 & 62.44 \\ \hline
2800 & 61.27 & 63.21 & 64.11 & 61.74 & 62.58 \\ \hline
3600 & 59.79 & 62.07 & 64.18 & 63.46 & 62.38 \\ \hline
\textbf{4000} & \textbf{60.68} & \textbf{63.28} & \textbf{65.71} & \textbf{64.1} & \textbf{63.44} \\ \hline
4500 & 60.81 & 63.55 & 65.19 & 62.76 & 63.08 \\ \hline
\end{tabular}
\end{center}
\caption{Validation accuracies for the LSTM models.}
\label{tab:lstm_val_perf}
\end{table}
\endgroup

\begingroup
\renewcommand*{\arraystretch}{1}
\begin{table}
\begin{center}
\begin{tabular}{|c|c|c|c|c|c|c|}
 \hline
 \multicolumn{7}{|c|}{DenseNet-style convolutional model} \\
 \cline{1-7}
 & Normalization & Fold 1 & Fold 2 & Fold 3 & Fold 4 & Average \\
  \hline
 Input embedding & IN & \textbf{64.82} & \textbf{65.91} & \textbf{69.14} & \textbf{66.65} & \textbf{66.63} \\ \hline
 Input embedding & BN & 62.34 & 65.72 & 66.07 & 63.98 & 64.53 \\ \hline
 No input embedding & IN & 62.71 & 63.64 & 65.69 & 61.19 & 63.30 \\ \hline
 No input embedding & BN & 57.95 & 60.97 & 63.47 & 58.39 & 60.19 \\ \hline
 \multicolumn{7}{|c|}{ResNet-style convolutional model} \\
 \cline{1-7}
  & Normalization & Fold 1 & Fold 2 & Fold 3 & Fold 4 & Average \\
  \hline
 Input embedding & IN & \textbf{65.59} & \textbf{68.94} & \textbf{69.45} & \textbf{67.09} & \textbf{67.76} \\ \hline
 Input embedding & BN & 62.49 & 65.57 & 65.65 & 64.57 & 64.57 \\ \hline
 No input embedding & IN & 61.57 & 59.15 & 62.21 & 63.12 & 61.51 \\ \hline
 No input embedding & BN & 58.75 & 61.22 & 61.90 & 58.47 & 60.09 \\ \hline
\end{tabular}
\end{center}
\caption{Validation accuracies for DenseNet-style convolutional models (IN = Instance normalization, BN =  Batch normalization)}
\label{tab:conv_val_perf}
\end{table}


\endgroup



\section{Convolutional Architectures} \label{sec:app_schematic_conv}
\begin{figure}[t]
    \centering 
    \includegraphics[width=\linewidth]{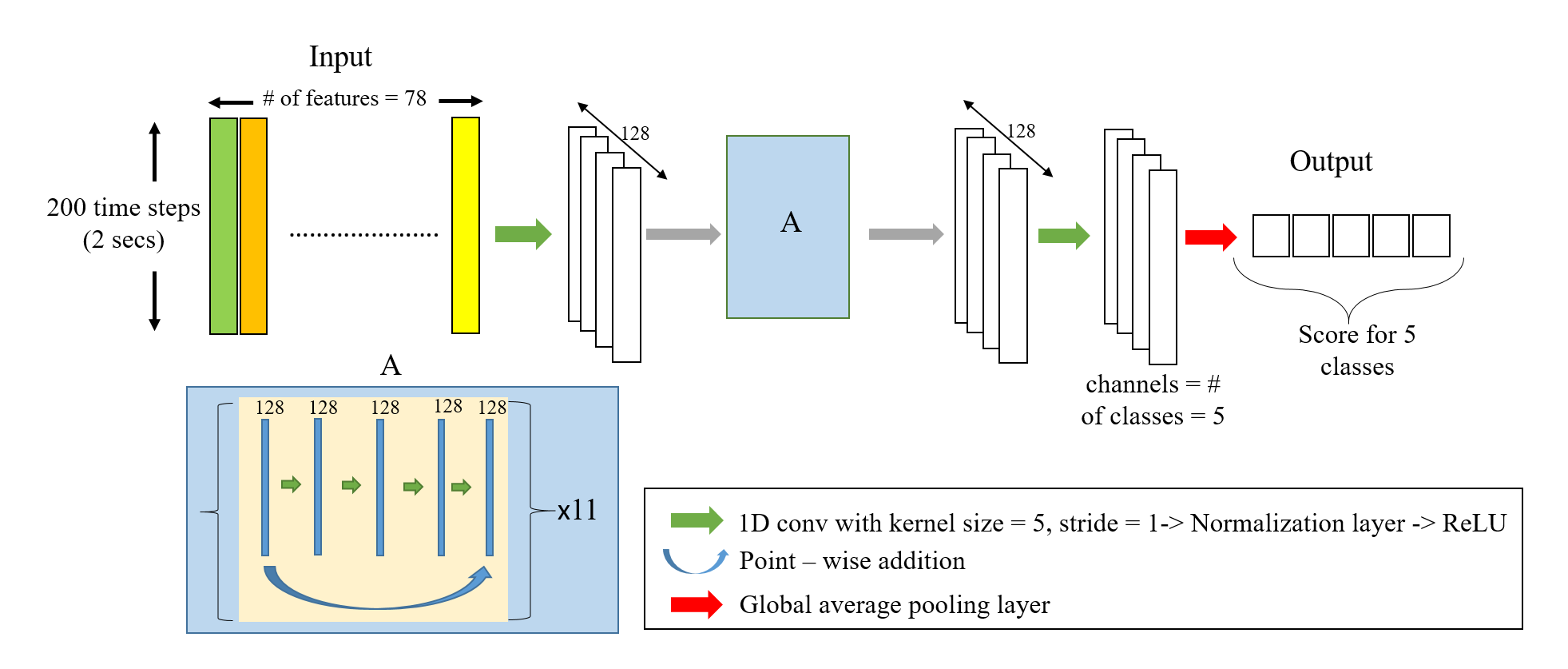} 
    \caption{Diagram of the ResNet-style convolutional network}
    \label{fig:res_net_trad} 
\end{figure}

\begin{figure}[t]
    \centering 
    \includegraphics[width=\linewidth]{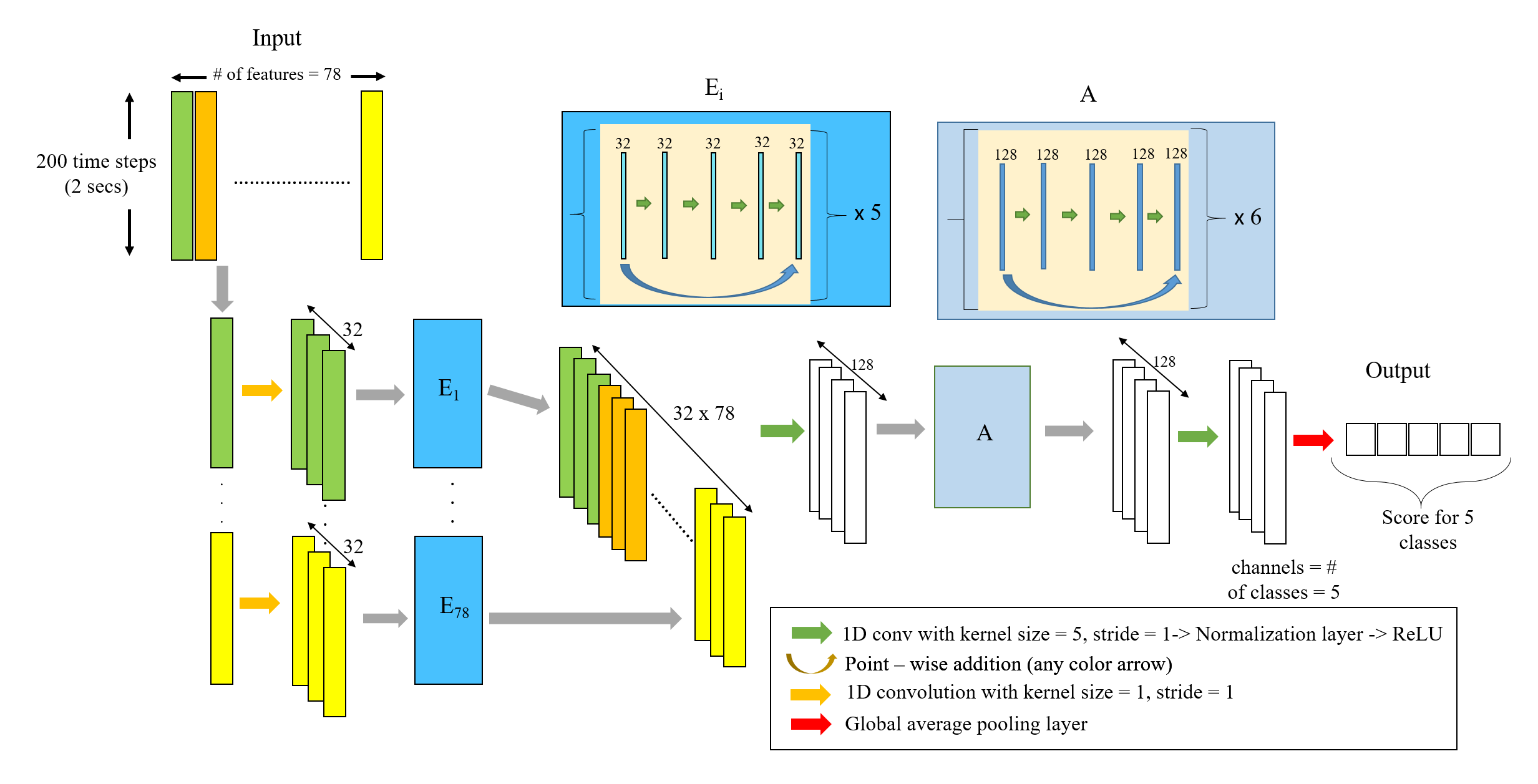} 
    \caption{Diagram of the ResNet-style convolutional network incorporating the input-embedding module described in Section~\ref{sec:input_embeddings}.}
    \label{fig:res_net_novel} 
\end{figure}

\begin{figure}[t]
    \centering 
    \includegraphics[width=\linewidth]{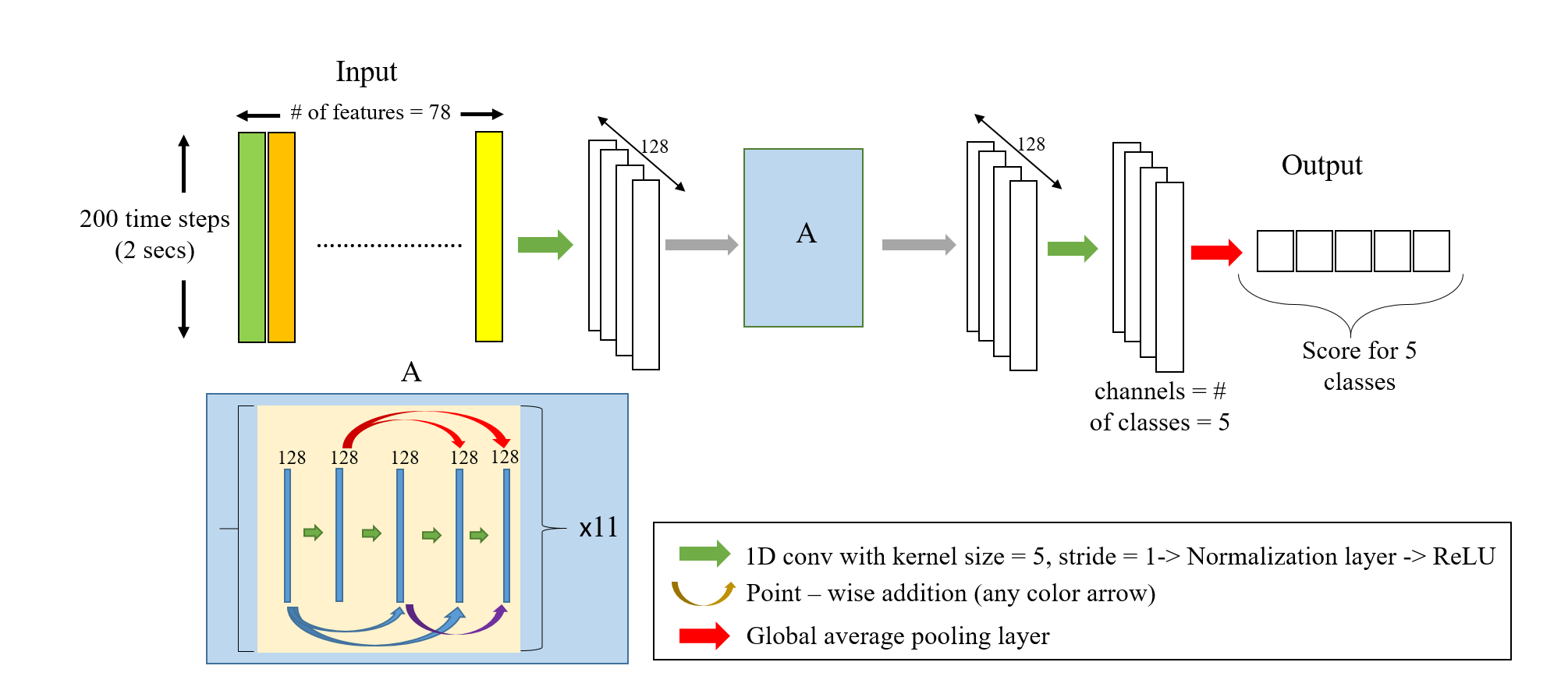} 
    \caption{Diagram of the DenseNet-style convolutional network}
    \label{fig:dense_net_trad} 
\end{figure}

\begin{figure}[t]
    \centering 
    \includegraphics[width=\linewidth]{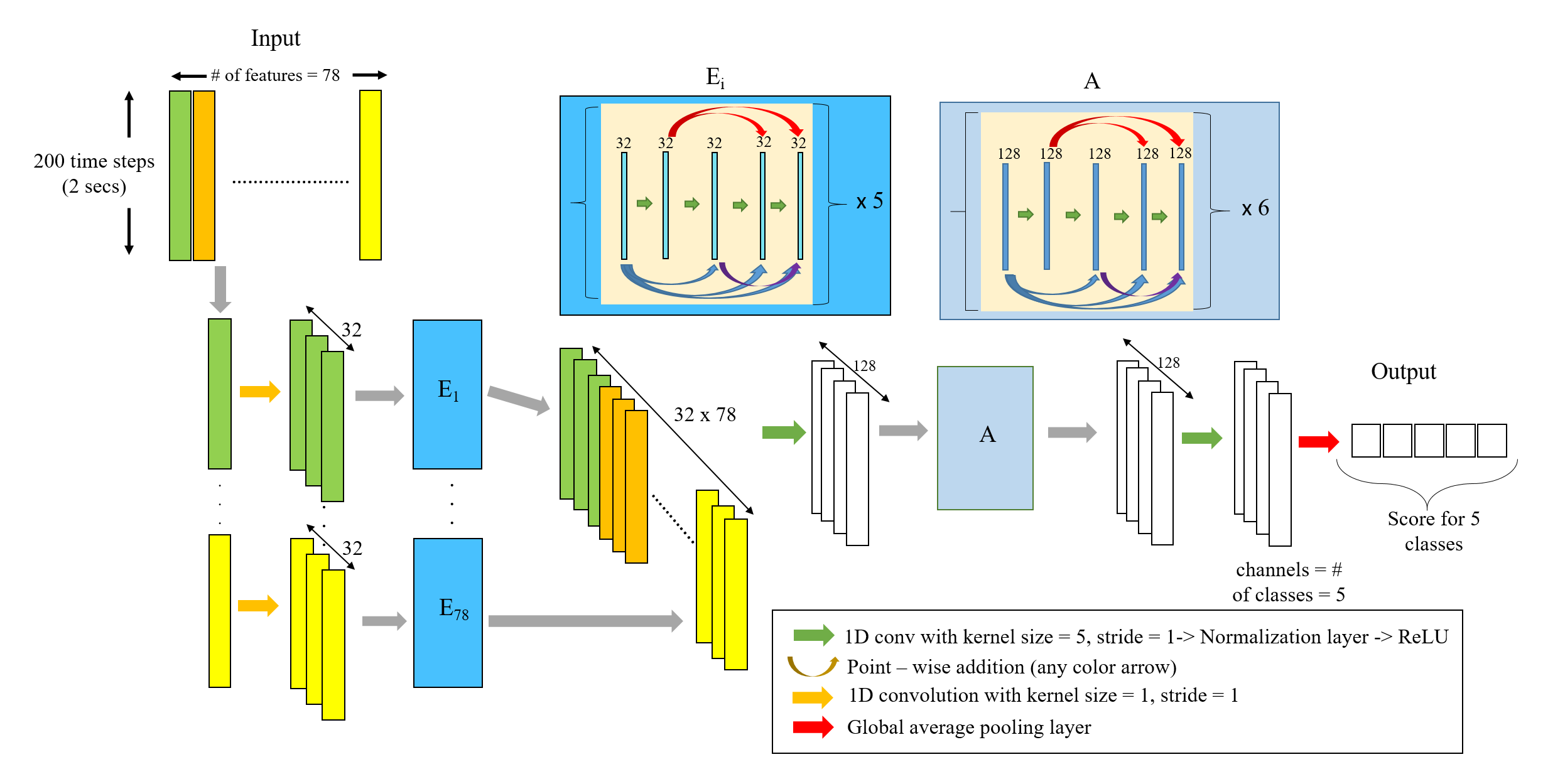} 
    \caption{Diagram of the DenseNet-style convolutional network incorporating the input-embedding module described in Section~\ref{sec:input_embeddings}.}
    \label{fig:dense_net_novel}  
\end{figure}

Figures~\ref{fig:res_net_trad} and \ref{fig:dense_net_trad} provide a detailed description of the baseline convolutional neural networks used for our experiments. Figure~\ref{fig:res_net_novel},  and  \ref{fig:dense_net_novel} show the modified architectures, which incorporate the input-embedding module described in Section~\ref{sec:input_embeddings}.


\end{document}